\documentclass[11pt]{article}
\usepackage{relsize}
\usepackage[top=1cm, bottom=1cm, left=1cm, right=1cm]{geometry}
\usepackage{mathrsfs}
\usepackage[fleqn]{amsmath}
\usepackage{amsfonts}
\usepackage{cite}
\usepackage{feynmf}
\usepackage[titletoc]{appendix}
\usepackage{graphicx}
\usepackage{indentfirst}
\usepackage{epstopdf}

\title{\Large{Nonperturbative solutions of Dyson-Schwinger equations in QED$_3$}}
\date{}
\begin{document}

\author{Wei Wei$^{1}$, Pei-lin Yin$^{1}$, and Hong-shi Zong$^{1,2,3}$\footnote{Email:zonghs@nju.edu.cn}\\
\small 1 Department of Physics, Nanjing University, Nanjing 210093, China\\
\small 2 Joint Center for Particle, Nuclear Physics and Cosmology, Nanjing 210093, China\\
\small 3 State Key Laboratory of Theoretical Physics, Institute of Theoretical Physics, CAS, Beijing, 100190, China}

\unitlength = 1mm
\maketitle
\abstract{The studies of Dyson-Schwinger Equations (DSEs) provide us with insights into nonperturbative phenomenon of quantum field theory. However, DSEs are essentially an infinite set of coupled Green's functions, it's necessary to decouple parts of the equations which are thought of major physical importance to make the solution of these equations possible. Although the results are model-dependent, no qualitative deviations from exact solutions are expected with properly chosen truncation scheme. In this article, a globally convergent numerical method for the solution of the DSEs of QED$_3$ in Euclidean space is presented. This method can be adapted for more complex problems, however, it also shows its limitations when adopted in problems such as the searching for Wigner solutions.
%
% Introduction
%
\section{Introduction}
Quantum Electrodynamics in (2+1) dimensions (QED$_3$) is studied for a variety of reasons~\cite{q0,q1,q2,q3,ql,qf,qy,q4,q5,q6}. It possesses nonperturbative features such as confinement and dynamical mass generation while avoids additional complication of non-Abelian theory\cite{ToyModel}. Furthermore, it has also been proposed that QED$_3$ can be employed to study the High-$T_c$ cuprate superconductors, where the motion of electrons are confined to two-dimensional planes\cite{Superconductor1, Superconductor2, Superconductor3, Superconductor 4}. However, as the analysis of nonperturbative solution cannot be conducted on an order-by-order basis from certain starting point, the solution of DSEs relies on self-consistent physically motivated models\cite{ToyModel}.

Although the normal fixed-point iteration works well in the calculation DSEs with bare vertex and the minimal Ball-Chiu (BC1) vertex~\cite{bc1}, this method can't provide sufficient sampling at the the singular point when calculating the dressing function using the full Ball-Chiu (BC) vertex~\cite{fbc}. This problem can be resolved by adopting the polynomial expansion of of the unknown function\cite{Thesis} and apply Newton Method to the iteration of the coefficients. In this method, as long as coefficients of the expansion are known, we can calculate the unknown function over the range of expansion. The infrared and ultraviolet behavior can be determined analytically. However, the local Newton depend critically on the initial values of iteration, therefore a algorithm with global convergence is adopted to reduce the dependence on the initial value of iterations.
%
% Formulation of Dyson-Schwinger Equations in QED$_3$
%
\section{Formulation of Dyson-Schwinger equations in QED$_3$}
\subsection{functional formulation of QED$_3$}
The Lagrangian density of QED$_3$ with $N_f$ fermion flavors can be written as
\begin{equation}
\mathscr{L}(\bar{\psi},\psi, A_{\mu}) = \sum^{N_f}_{f=1}\bar{\psi}^{f}(i\partial \!\!\!/-m^f_0 + e^f_0 A\!\!\!/)\psi^f - \frac{1}{4}F_{\mu\nu}F^{\mu\nu} - \frac{1}{2\xi}(\partial_{\mu}A^{\mu})^2,
\end{equation}
where $f$ is the flavor index, $N_f$ is the total number of flavors and $\xi$ is the gauge-fixing constant.

The generating functional of correlation functions defined via this Lagrangian density reads:
\begin{equation}
Z[\eta,\bar{\eta},J_{\mu}] = \int [d\psi^f(x)][d\bar{\psi^f(x)}]d[A_{\mu}(x)]exp\left\{i\int d^3 x [\mathscr{L}(\bar{\psi},\psi, A_{\mu}) + \sum_{f}(\bar{\psi^f}\eta^f + \bar{\eta^f}\psi^f) + A_{\mu}J^{\mu}]\right\}.
\end{equation}

The generating functional of connected Green's functions $G_c[\eta,\bar{\eta},J_{\mu}]$ is related to $Z[\eta,\bar{\eta},J_{\mu}]$ by \cite{ConnectedGraph}
\begin{equation}
Z[\eta,\bar{\eta},J_{\mu}] = exp(G_c[\eta,\bar{\eta},J_{\mu}]).
\end{equation}

We can  perform the Legendre transformation of $G_c[\eta,\bar{\eta},J_{\mu}]$ to obtain the generating functional of proper vertices
\begin{equation}
i\Gamma[\psi,\bar{\psi},A_{\mu}] = G_c[\eta,\bar{\eta},J_{\mu}] - i\int d^3\,x[\bar{\psi}^f\eta^f + \bar{\eta}^f\psi^f + A_{\mu}J^{\mu}],
\end{equation}
with identifications
\begin{equation}
\psi^f = \frac{\delta G_c}{i\delta\bar{\eta}^f},\ \ \ \  \bar{\psi}^f = -\frac{\delta G_c}{i\delta\eta^f},\ \ \ \   A_{\nu} = \frac{\delta G_c}{i\delta J^{\nu}}.
\end{equation}

It can be shown that $\Gamma[\psi,\bar{\psi},A_{\mu}]$ generates the 1-PI Green's functions\cite{ConnectedGraph}. Then the fermion propagator $S(x,y)$, photon propagator $D^{\mu\nu}$ and and the 1-PI fermion-photon vertex can be expressed as
\begin{equation}
iS^f(x,y) = \frac{\delta^2 G_c}{i\delta \bar{\eta}^f(x)\,i\delta\eta^f(y)}\Bigg|_{\eta=0, \bar{\eta}=0, J = 0},
\end{equation}
\begin{equation}
iD_{\mu\nu}(x,y) = \frac{\delta^2G_c}{i\delta J^{\mu}(x)\,i\delta J^{\nu}(y)}\Bigg|_{\eta=0, \bar{\eta}=0, J = 0},
\end{equation}
\begin{equation}
e_0^f\Gamma(x,y;z)^f = \frac{\delta^3\Gamma}{\delta A_{\mu}(z)\delta\bar{\psi}(x)\delta\psi(y)}\Bigg|_{\eta=0, \bar{\eta}=0, J = 0}.
\end{equation}

 For simplicity, the flavor index will be omitted in the following context.
\subsection{Dyson-Schwinger equations for the fermion and photon propagators}

The complete two-point Green's function $S$ can be constructed out of the electron self-energy $\Sigma_{p}$ by
\begin{equation}
iS(p)=iS_0(p) + iS_0(p)[-i\Sigma(p)]iS_0(p) + iS_0(p)[-i\Sigma(p)]iS_0(p)[-i\Sigma(p)]iS_0(p) + \cdot\ \cdot \ \cdot
\end{equation}
where
\begin{equation}
S_0(p) = \frac{1}{p\!\!\!/-m_0},
\end{equation}
then we have
\begin{equation}
S(p) = \frac{1}{p\!\!\!/-m_0-\Sigma(p)}.
\end{equation}
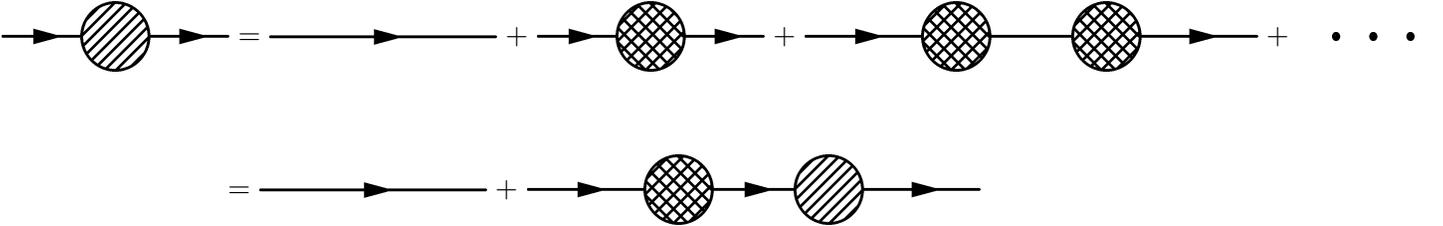
\begin{figure}[h]
\centering
\begin{fmffile}{40261}
\parbox{30mm}{
\begin{fmfgraph*}(30,20)
\fmfleft{i}
\fmfright{o}
\fmfv{d.sh=circle,d.f=shaded,d.si=.3w,label=$iS$, label.angle= -90, label.dist=0.25w}{v}
\fmf{fermion}{i,v,o}
\end{fmfgraph*}
}
$=$
\parbox{30mm}{
\begin{fmfgraph*}(30,20)
\fmfleft{i}
\fmfright{o}
\fmf{fermion,label=$iS_0$, label.angle= -90, label.dist=0.25w}{i,o}
\end{fmfgraph*}
}
$+$
\parbox{30mm}{
\begin{fmfgraph*}(30,20)
\fmfleft{i}
\fmfright{o}
\fmfv{d.sh=circle,d.f=hatched,d.si=.3w, label=$-i\Sigma$, label.angle= -100, label.dist=0.25w}{v}
\fmf{fermion,label=$iS_0$, label.angle= -90, label.dist=0.25w}{i,v,o}
\end{fmfgraph*}
}
$+$
\parbox{60mm}{
\begin{fmfgraph*}(60,20)
\fmfleft{i}
\fmfright{o}
\fmfv{d.sh=circle,d.f=hatched,d.si=.15w, label=$-i\Sigma$, label.angle= -100, label.dist=0.13w}{v1}
\fmfv{d.sh=circle,d.f=hatched,d.si=.15w, label=$-i\Sigma$, label.angle= -100, label.dist=0.13w}{v2}
\fmf{fermion,label=$iS_0$, label.angle= -90, label.dist=0.13w}{i,v1}
\fmf{plain,label=$iS_0$, label.angle= -90, label.dist=0.13w}{v1,v2}
\fmf{fermion,label=$iS_0$, label.angle= -90, label.dist=0.13w}{v2,o}
\end{fmfgraph*}
}
$+$
\parbox{20mm}{
\begin{fmfgraph*}(20,20)
\fmfleft{i}
\fmfv{d.sh=circle,d.f=hatched,d.si=thick}{v1}
\fmfv{d.sh=circle,d.f=hatched,d.si=thick}{v2}
\fmfv{d.sh=circle,d.f=hatched,d.si=thick}{v3}
\fmfright{o}
\fmf{phantom}{i,v1,v2,v3,o}
\end{fmfgraph*}
}
\end{fmffile}
%\end{figure}
%\begin{figure}[h]
%\centering
\begin{fmffile}{f4029}
%\parbox{30mm}{
%\begin{fmfgraph*}(30,20)
%\fmfleft{i}
%\fmfright{o}
%\fmf{phantom}{i,o}
%\end{fmfgraph*}
%}
$=$
\parbox{30mm}{
\begin{fmfgraph*}(30,20)
\fmfleft{i}
\fmfright{o}
\fmf{fermion,label=$iS_0$, label.angle= -90, label.dist=0.25w}{i,o}
%\fmfv{d.sh=circle,d.f=full,d.si=2thick, label=$iS_0$, label.angle= -90, label.dist=0.125w}{v}
\end{fmfgraph*}
}
$+$
\parbox{60mm}{
\begin{fmfgraph*}(60,20)
\fmfleft{i}
\fmfright{o}
%\fmfv{d.sh=circle,d.f=full,d.si=thin, label=$iS_0$, label.angle= -90, label.dist=0.125w}{v1}
\fmfv{d.sh=circle,d.f=hatched,d.si=.15w, label=$-i\Sigma$, label.angle= -100, label.dist=0.12w}{v2}
\fmfv{d.sh=circle,d.f=shaded,d.si=.15w,label=$iS$, label.angle= -90, label.dist=0.12w}{v3}
\fmf{fermion,label=$iS_0$,label.angle=-90,label.dist=0.12w}{i,v2}
\fmf{fermion}{v2,v3,o}
\end{fmfgraph*}
}
\parbox{30mm}{
\begin{fmfgraph*}(30,20)
\fmfleft{i}
\fmfright{o}
\fmf{phantom}{i,o}
\end{fmfgraph*}
}
\end{fmffile}
\caption{The complete fermion propagator as a sum of 1PI self-energy insertions.}
\end{figure}

The Schwinger-Dyson equation for the fermion self-energy is given by
\begin{equation} \label{Sigma}
-i\Sigma(p) = -\int \frac{d^3 k}{(2\pi)^3}\gamma_{\mu}iS(k)\Gamma_{\nu}(k,p)iD_{\mu\nu}(q),
\end{equation}
where $q = p - k$.
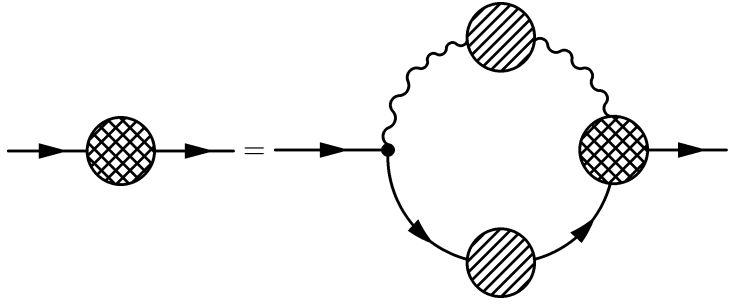
\begin{figure}[h]
\centering
\begin{fmffile}{261}
\parbox{30mm}{
\begin{fmfgraph*}(30,40)
\fmfleft{i}
\fmfright{o}
\fmfv{d.sh=circle,d.f=hatched,d.si=.3w}{v}
\fmf{fermion,label=$p$}{i,v,o}
\end{fmfgraph*}
}
$=$
\parbox{60mm}{
\begin{fmfgraph*}(60,30)
\fmfleft{i}
\fmfright{o}
\fmf{fermion,label=$p$}{i,v3}
\fmf{fermion,label=$p$}{v1,o}
\begin{fmfsubgraph}(0.25w,0)(0.5w,1.0h)
\fmfsurroundn{v}{4}
\fmfcyclen{phantom, right=0.45,tag=1}{v}{4}
\fmfcyclen{phantom, right=0.45,tag=2}{v}{4}
\fmfcyclen{phantom, right=0.35,tag=3}{v}{4}
\fmfcyclen{phantom, right=0.35,tag=4}{v}{4}
\fmfv{d.sh=circle,d.f=full,d.si=2thick,label=$\mu$,label.angle=0}{v3}
\fmfblob{0.15w}{v2,v4}
\fmfv{d.sh=circle,d.f=hatched,d.si=.15w,label=$\nu$,label.angle=180,label.dist=.1w}{v1}
\fmfposition
\fmfipath{p[]}
\fmfiset{p1}{vpath1(__v1,__v2)}
\fmfiset{p2}{vpath2(__v2,__v3)}
\fmfiset{p3}{vpath2(__v3,__v4)}
\fmfiset{p4}{vpath2(__v4,__v1)}
\fmfi{photon,label=$p-k$}{subpath (0,length(p1)) of p1}
\fmfi{photon,label=$p-k$}{subpath (0,length(p2)) of p2}
\fmfi{fermion,label=$k$}{subpath (0,length(p3)) of p3}
\fmfi{fermion,label=$k$}{subpath (0,length(p4)) of p4}
\end{fmfsubgraph}
\end{fmfgraph*}
}
\end{fmffile}

\caption{Diagrammatic representation of the integral equation for fermion self-energy $\Sigma(p)$.}
\end{figure}

Similarly, the complete photon propagator can be written in terms of self-energy $\Pi_{\mu\nu}(q)$ as
\begin{equation}
iD^{\mu\nu}(q) = iD_0^{\mu\nu}(q) + iD_0^{\mu\sigma}(q)[i\Pi_{\sigma\lambda}(q)]iD_0^{\lambda\nu}(q) + \cdot\ \cdot \ \cdot
\end{equation}
where
\begin{equation}
D_0^{\mu\nu}(q)=\frac{-g^{\mu\nu}}{q^2}+(1-\xi)\frac{q^{\mu}q^{\nu}}{q^2},
\end{equation}
we obtain
\begin{equation}
D^{\mu\nu}(q)=D_0^{\mu\nu}(q)-D_0^{\mu\sigma}(q)\Pi_{\sigma\lambda}(q)D^{\lambda\nu}(q)+ \cdot\ \cdot\ \cdot
\end{equation}
\begin{figure}[h]
\centering
\begin{fmffile}{50051}
\parbox{30mm}{
\begin{fmfgraph*}(30,20)
\fmfleft{i}
\fmfright{o}
\fmfv{d.sh=circle,d.f=shaded,d.si=.3w,label=$iD$, label.angle= -90, label.dist=0.25w}{v}
\fmf{photon}{i,v,o}
\end{fmfgraph*}
}
$=$
\parbox{30mm}{
\begin{fmfgraph*}(30,20)
\fmfleft{i}
\fmfright{o}
\fmf{photon,label=$iD_0$, label.angle= -90, label.dist=0.25w}{i,o}
\end{fmfgraph*}
}
$+$
\parbox{30mm}{
\begin{fmfgraph*}(30,20)
\fmfleft{i}
\fmfright{o}
\fmfv{d.sh=circle,d.f=hatched,d.si=.3w, label=$i\Pi$, label.angle= -90, label.dist=0.25w}{v}
\fmf{photon,label=$iD_0$, label.angle= -90, label.dist=0.25w}{i,v,o}
\end{fmfgraph*}
}
$+$
\parbox{60mm}{
\begin{fmfgraph*}(60,20)
\fmfleft{i}
\fmfright{o}
\fmfv{d.sh=circle,d.f=hatched,d.si=.15w, label=$i\Pi$, label.angle= -90, label.dist=0.13w}{v1}
\fmfv{d.sh=circle,d.f=hatched,d.si=.15w, label=$i\Pi$, label.angle= -90, label.dist=0.13w}{v2}
\fmf{photon,label=$iD_0$, label.angle= -90, label.dist=0.13w}{i,v1}
\fmf{photon,label=$iD_0$, label.angle= -90, label.dist=0.13w}{v1,v2}
\fmf{photon,label=$iD_0$, label.angle= -90, label.dist=0.13w}{v2,o}
\end{fmfgraph*}
}
$+$
\parbox{20mm}{
\begin{fmfgraph*}(20,20)
\fmfleft{i}
\fmfv{d.sh=circle,d.f=hatched,d.si=thick}{v1}
\fmfv{d.sh=circle,d.f=hatched,d.si=thick}{v2}
\fmfv{d.sh=circle,d.f=hatched,d.si=thick}{v3}
\fmfright{o}
\fmf{phantom}{i,v1,v2,v3,o}
\end{fmfgraph*}
}
\end{fmffile}
%\end{figure}
%\begin{figure}[h]
%\centering
\begin{fmffile}{f5004}
%\parbox{30mm}{
%\begin{fmfgraph*}(30,20)
%\fmfleft{i}
%\fmfright{o}
%\fmf{phantom}{i,o}
%\end{fmfgraph*}
%}
$=$
\parbox{30mm}{
\begin{fmfgraph*}(30,20)
\fmfleft{i}
\fmfright{o}
\fmf{photon,label=$iD_0$, label.angle= -90, label.dist=0.25w}{i,o}
%\fmfv{d.sh=circle,d.f=full,d.si=2thick, label=$iS_0$, label.angle= -90, label.dist=0.125w}{v}
\end{fmfgraph*}
}
$+$
\parbox{60mm}{
\begin{fmfgraph*}(60,20)
\fmfleft{i}
\fmfright{o}
%\fmfv{d.sh=circle,d.f=full,d.si=thin, label=$iS_0$, label.angle= -90, label.dist=0.125w}{v1}
\fmfv{d.sh=circle,d.f=hatched,d.si=.15w, label=$i\Pi$, label.angle= -90, label.dist=0.12w}{v2}
\fmfv{d.sh=circle,d.f=shaded,d.si=.15w,label=$iD$, label.angle= -90, label.dist=0.12w}{v3}
\fmf{photon,label=$iD_0$,label.angle=-90,label.dist=0.12w}{i,v2}
\fmf{photon}{v2,v3,o}
\end{fmfgraph*}
}
\parbox{30mm}{
\begin{fmfgraph*}(30,20)
\fmfleft{i}
\fmfright{o}
\fmf{phantom}{i,o}
\end{fmfgraph*}
}
\end{fmffile}
\caption{The complete photon propagator as a sum of proper 1PI photon self-energy.}
\end{figure}
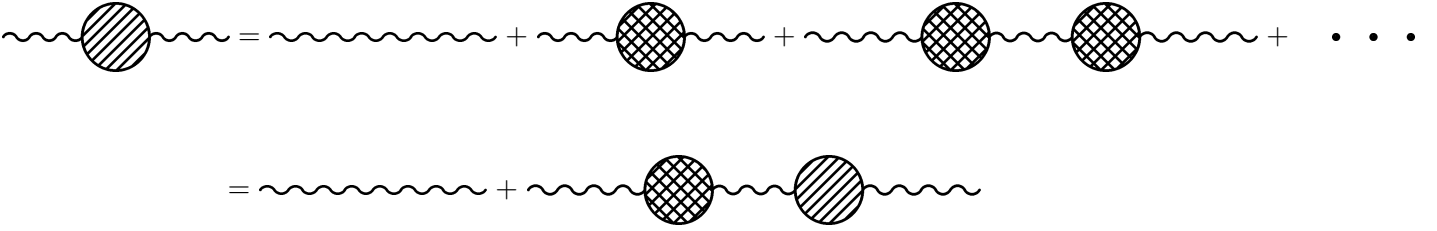

%On the other hand, based on the Lorentz structure analysis, the dressed fermion propagator has the general form
%\begin{equation} \label{GeneralForm}
%S^{-1}(p) = i p\!\!\!/ A(p^2) + B(p^2),
%\end{equation}
%Substitute Eq. \eqref{GeneralForm} into \eqref{Sigma}, we obtain the following set of couple integral equations
%\begin{equation} \label{A}
%A(p^2) = 1 - \frac{1}{4p^2}\int \frac{d^3 k}{(2\pi)^2}Tr[i(\gamma P)\gamma_{\sigma}S(k)\Gamma_{\nu}D_{\sigma\nu}(q)],
%\end{equation}
%\begin{equation} \label{B}
%B(p^2) = \frac{1}{4}\int \frac{d^3(k)}{(2\pi)^3}Tr[\gamma_{\sigma}S(k)\Gamma_{\nu}(p,k)D_{\sigma\nu}(q)],
%\end{equation}
\newpage
The DSEs for the photon self-energy reads
\begin{equation} \label{PiTensor}
i\Pi_{\mu\nu}(q) = N_fe_0^2\int \frac{d^3k}{(2\pi)^3}Tr[iS(k)\gamma_{\mu}iS(k+q)\Gamma_{\nu}(k+q,k)].
\end{equation}
\begin{figure}[h]
\centering
\begin{fmffile}{266}
\parbox{30mm}{
\begin{fmfgraph*}(30,40)
\fmfleft{i}
\fmfright{o}
\fmfv{d.sh=circle,d.f=hatched,d.si=.3w}{v}
\fmf{photon,label=$q$}{i,v,o}
\end{fmfgraph*}
}
$=$
\parbox{60mm}{
\begin{fmfgraph*}(60,30)
\fmfleft{i}
\fmfright{o}
\fmf{photon,label=$q$}{i,v3}
\fmf{photon,label=$q$}{v1,o}
\begin{fmfsubgraph}(0.25w,0)(0.5w,1.0h)
\fmfsurroundn{v}{4}
\fmfcyclen{phantom, right=0.45,tag=1}{v}{4}
\fmfcyclen{phantom, right=0.45,tag=2}{v}{4}
\fmfcyclen{phantom, right=0.35,tag=3}{v}{4}
\fmfcyclen{phantom, right=0.35,tag=4}{v}{4}
\fmfv{d.sh=circle,d.f=full,d.si=2thick,label=$\mu$,label.angle=0}{v3}
\fmfblob{0.15w}{v2,v4}
\fmfv{d.sh=circle,d.f=hatched,d.si=.15w,label=$\nu$,label.angle=180,label.dist=.1w}{v1}
\fmfposition
\fmfipath{p[]}
\fmfiset{p1}{vpath1(__v1,__v2)}
\fmfiset{p2}{vpath2(__v2,__v3)}
\fmfiset{p3}{vpath2(__v3,__v4)}
\fmfiset{p4}{vpath2(__v4,__v1)}
\fmfi{fermion,label=$k$}{subpath (0,length(p1)) of p1}
\fmfi{fermion,label=$k$}{subpath (0,length(p2)) of p2}
\fmfi{fermion,label=$k+q$}{subpath (0,length(p3)) of p3}
\fmfi{fermion,label=$k+q$}{subpath (0,length(p4)) of p4}
\end{fmfsubgraph}
\end{fmfgraph*}
}
\end{fmffile}
\caption{Diagrammatic representation of the integral equation for $\Pi_{\mu\nu}(q)$.}
\end{figure}
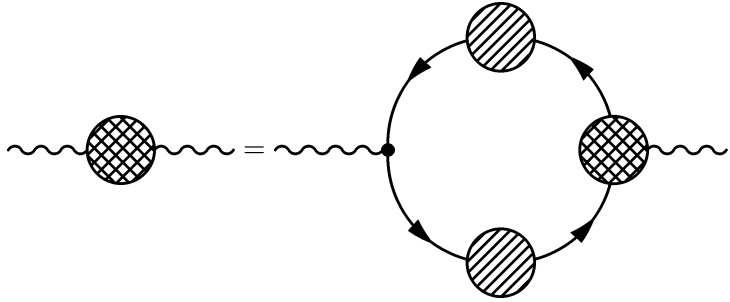
%where $D^{0}_{\mu\nu} = (\delta_{\mu\nu}-q_{\mu}q_{\nu}/q^2)$ is the bare propagator in Landau gauge and $\Pi(q)$ is the vacuum polarization.

%The Ward-Takahashi identity restricts $\Pi(q)$ to the following form
%\begin{equation} \label{PiScalar}
%\Pi(q) = (q^2\delta_{\mu\nu} - q_{\mu}q_{\nu})\Pi(q^2)
%\end{equation}
%The ultraviolet divergence in $\Pi(q^2)$ can be projected out by the following operator\cite{Projector}
%\begin{equation}
%\mathcal{P}_{\mu\nu} = \delta_{\mu\nu} - 3\frac{q_{\mu}}{q_{\nu}}{q^2}
%\end{equation}

\subsection{Dyson-Schwinger equation for the fermion-photon vertex}

The Dyson-Schwinger equation for fermion-photon vertex can be shown to be \cite{Bjorken}
\begin{equation}
\Gamma_{\mu}(p^{\prime},p) = \gamma_{\mu} + \int \frac{d^3k}{(2\pi)^3}(iS(p^{\prime}+k)\Gamma_{\mu}(p^{\prime}+k,p+k))iS(p+k)K(p+k,p^{\prime}+k,k),
\end{equation}
where $K$ is the fermion-antifermion scattering kernel.

The field theory can be completely determined when all of its Green's functions are known. DSEs connected all these $n-$point functions through a set of infinitely coupled integral function. In the derivation for the fermion-photon vertex, we encounter another unknown 4-point function $K$ which satisfies its own integral equation and will be coupled to a 5-point Green's function.To make the DSEs tractable, it necessary to introduce the truncation at certain $n$. We can choose $S, D_{\mu\nu}, \Gamma_{\mu}$ as the building block of our theory. Quantities such as $K$ be expressed in terms of $S, D_{\mu\nu}, \Gamma_{\mu}$ by the analysis of their topological structure. The theory can then be determined self-consistently in this way. To compensate for the strategy of truncation, we have to introduce physically motivated vertex in our calculation.Common restriction on vertex are various symmetries and the known asymptotic behaviors of dressing functions. Several attempts have been made to determine the form of $\Gamma_{\nu}$\cite{Vertex1, Vertex2}. The simplest vertex is the bare one, however, this violate the Ward-Takahashi identity which is a consequence of the gauge invariance of the theory. In this article, a numerical method for the solution of DSEs in the BC vertex which satisfied the Ward-Takahahsi identity is presented.
Due to the singular structure of the kernel of integration, the fixed point can't be to calculate the dressing function with precision to satisfactory degree. The Newton's iterative method is adopted in the literature \cite{Thesis}. However, the dressing functions are not calculated in the same loop in the method proposed by the author \cite{Thesis}. And the author has to manually cancel the infrared divergence which is not supposed to appear in the final result of calculation. The method presented in this article will solve the three coupled equations consistently.
\begin{figure}[h]
\centering
\begin{fmffile}{3064}
\parbox{30mm}{
\begin{fmfgraph*}(40,40)
\fmfleft{i}
\fmfright{o1,o2}
\fmf{photon}{i,v}
\fmf{fermion,label=$p$,label.side=left, label.dist=0.08w}{o1,v}
\fmf{fermion,label=$p^{\prime}$,label.side=left}{v,o2}
\fmfv{d.sh=circle,d.f=shaded,d.si=.2w, label=$\mu$, label.angle= -120, label.dist=0.15w}{v}
\end{fmfgraph*}
}
$\mathlarger{\mathlarger{\mathlarger{\mathlarger{=}}}}$\,\,
\parbox{40mm}{
\begin{fmfgraph*}(40,40)
\fmfleft{i}
\fmfright{o1,o2}
\fmfv{d.sh=circle,d.f=full,d.si=2thick, label=$\mu$, label.angle= -120, label.dist=0.05w}{v}
\fmf{photon}{i,v}
\fmf{fermion,label=$p$, label.dist=0.08w}{o1,v}
\fmf{fermion,label=$p^{\prime}$,label.side=left}{v,o2}
\end{fmfgraph*}
}
$\mathlarger{\mathlarger{\mathlarger{\mathlarger{+}}}}$\,\,
\parbox{100mm}{
\begin{fmfgraph*}(100,40)
\fmfleft{i}
\fmfv{d.sh=circle,d.f=shaded,d.si=.08w}{v1}
\fmfv{d.sh=circle,d.f=shaded,d.si=.08w}{v2}
\fmf{photon}{i,v}
\fmfright{o3,o4}
\fmf{fermion}{v,v1}
\fmf{fermion}{v2,v}
\fmfpolyn{empty, label=$K$, tension=0.8}{K}{4}
\fmf{fermion,label=$p^{\prime}+k$,label.side=left}{v1,K3}
\fmf{fermion,label=$p+k$, label.side=left}{K4,v2}
\fmf{fermion,label=$p$,label.side=left, label.dist=0.02w}{o3,K1}
\fmf{fermion,label=$p^{\prime}$,label.side=left, label.dist=0.02w}{K2,o4}
\fmffreeze
\fmfv{d.sh=circle,d.f=hatched,d.si=.08w, label=$\mu$, label.angle= -120, label.dist=0.06w}{v}
\fmfshift{0,0.08h}{v1}
\fmfshift{0,-0.08h}{v2}
\end{fmfgraph*}
}
\end{fmffile}
\end{figure}
\begin{figure}[h]
\centering
\begin{fmffile}{3066}
$\mathlarger{\mathlarger{\mathlarger{\mathlarger{=}}}}$\,\,
\parbox{40mm}{
\begin{fmfgraph*}(40,40)
\fmfleft{i}
\fmfright{o1,o2}
\fmfv{d.sh=circle,d.f=full,d.si=2thick, label=$\mu$, label.angle= -120, label.dist=0.05w}{v}
\fmf{photon}{i,v}
\fmf{fermion,label=$p$, label.dist=0.08w}{o1,v}
\fmf{fermion,label=$p^{\prime}$,label.side=left}{v,o2}
\end{fmfgraph*}
}
$\mathlarger{\mathlarger{\mathlarger{\mathlarger{+}}}}$\,\,
\parbox{100mm}{
\begin{fmfgraph*}(100,40)
\fmfleft{i}
\fmfv{d.sh=circle,d.f=shaded,d.si=.08w}{v1}
\fmfv{d.sh=circle,d.f=shaded,d.si=.08w}{v2}
\fmf{photon}{i,v}
\fmfright{o3,o4}
\fmf{fermion}{v,v1}
\fmf{fermion}{v2,v}
\fmfpolyn{empty, label=$K$, tension=0.8}{K}{4}
\fmf{fermion,label=$p^{\prime}+k$,label.side=left}{v1,K3}
\fmf{fermion,label=$p+k$, label.side=left}{K4,v2}
\fmf{fermion,label=$p$,label.side=left, label.dist=0.02w}{o3,K1}
\fmf{fermion,label=$p^{\prime}$,label.side=left, label.dist=0.02w}{K2,o4}
\fmffreeze
\fmfv{d.sh=circle,d.f=full,d.si=2thick, label=$\mu$, label.angle= -120, label.dist=0.02w}{v}
\fmfshift{0,0.08h}{v1}
\fmfshift{0,-0.08h}{v2}
\end{fmfgraph*}
}
\end{fmffile}
\end{figure}
\begin{figure}[h]
\centering
\begin{fmffile}{3087}
$\mathlarger{\mathlarger{\mathlarger{\mathlarger{+}}}}$\,\,
\parbox{100mm}{
\begin{fmfgraph*}(120,40)
\fmfleft{i}
\fmfv{d.sh=circle,d.f=shaded,d.si=.07w}{v1}
\fmfv{d.sh=circle,d.f=shaded,d.si=.07w}{v2}
\fmf{photon}{i,v}
\fmfright{o3,o4}
\fmf{fermion}{v,v1}
\fmf{fermion}{v2,v}
\fmfpolyn{empty, label=$K$, tension=0.4}{K}{4}
\fmf{fermion,label=$p^{\prime}+k_1$,label.side=left,label.dist=0.05w}{v1,K3}
\fmf{fermion,label=$p+k_1$, label.side=left,label.dist=0.05w}{K4,v2}
\fmf{fermion}{vv2,K1}
\fmf{fermion}{K2,vv1}
\fmfv{d.sh=circle,d.f=full,d.si=2thick, label=$\mu$, label.angle= -120, label.dist=0.02w}{v}
\fmfv{d.sh=circle,d.f=shaded,d.si=.07w}{vv1}
\fmfv{d.sh=circle,d.f=shaded,d.si=.07w}{vv2}
\fmfpolyn{empty, label=$K$, tension=0.75}{KK}{4}
\fmf{fermion,label=$p^{\prime}+k_2$,label.side=left,label.dist=0.05w}{vv1,KK3}
\fmf{fermion,label=$p+k_2$, label.side=left,label.dist=0.05w}{KK4,vv2}
\fmf{fermion,label=$p$,label.side=left, label.dist=0.02w}{o3,KK1}
\fmf{fermion,label=$p^{\prime}$,label.side=left, label.dist=0.02w}{KK2,o4}
\fmffreeze
\fmfv{d.sh=circle,d.f=full,d.si=2thick, label=$\mu$, label.angle= -120, label.dist=0.02w}{v}
\fmfshift{0,0.08h}{v1}
\fmfshift{0,-0.08h}{v2}
\fmfshift{0,0.01h}{vv1}
\fmfshift{0,-0.01h}{vv2}
\end{fmfgraph*}
}
$\mathlarger{\mathlarger{\mathlarger{\mathlarger{+}}}}$
\parbox{20mm}{
\begin{fmfgraph*}(20,10)
\fmfleft{i}
\fmfv{d.sh=circle,d.f=hatched,d.si=thick}{v1}
\fmfv{d.sh=circle,d.f=hatched,d.si=thick}{v2}
\fmfv{d.sh=circle,d.f=hatched,d.si=thick}{v3}
\fmfright{o}
\fmf{phantom}{i,v1,v2,v3,o}
\end{fmfgraph*}
}
\end{fmffile}
\caption{Diagrammatic notation for the Dyson-Schwinger equation of fermion-photon vertex.}
\end{figure}
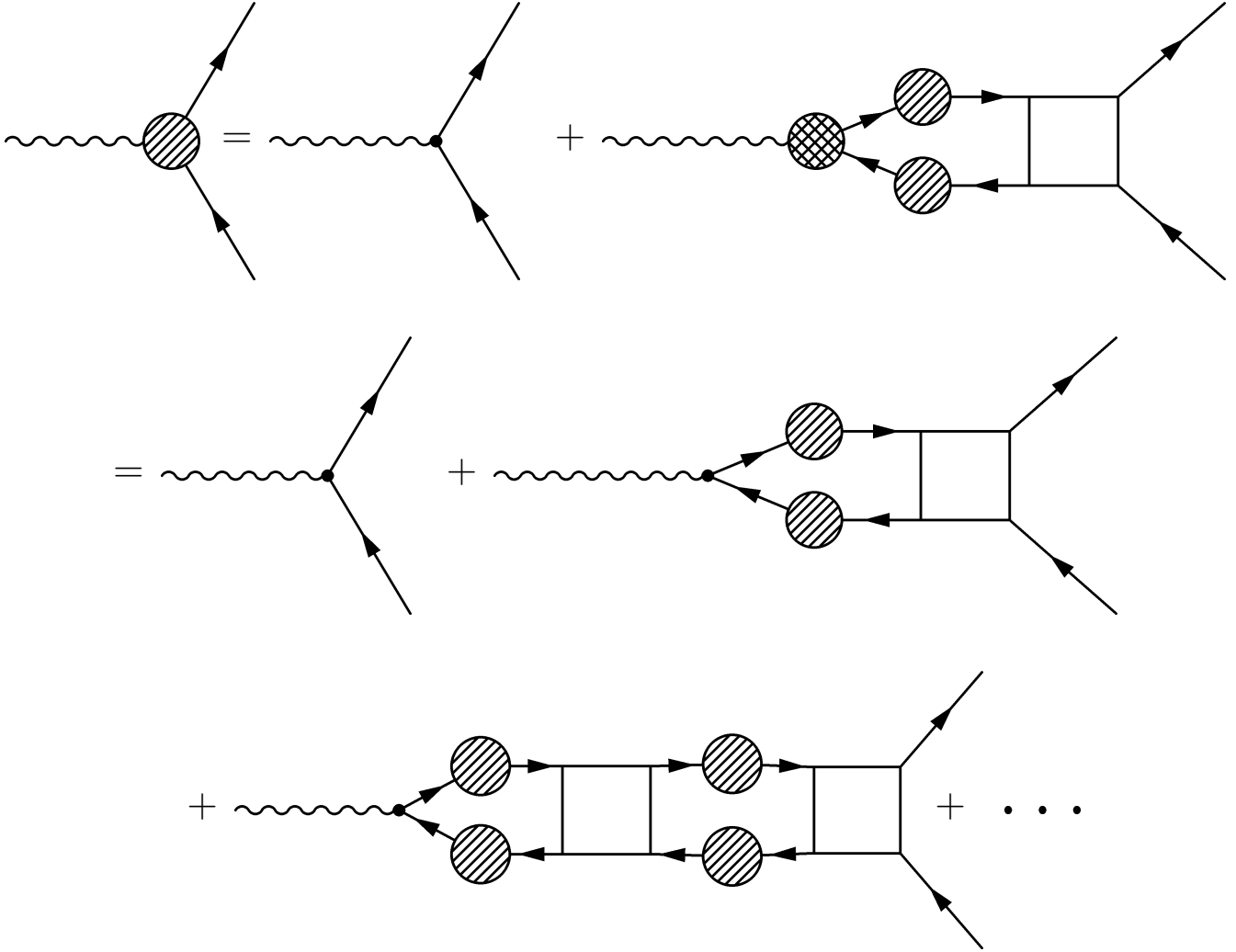
\subsection{The truncation scheme}
Following the construction in \cite{Vertex1}, it can be shown that the BC vertex takes the following form
\begin{equation}
\Gamma_{\nu}(p,k) = \frac{A(p^{2}+A(k^2))}{2}\gamma_{\nu} + \frac{(A(p^{2})-A(k^2))}{2(p^{2}-k^{2})}(\gamma \cdot p + \gamma \cdot k)(p+k)_{\nu} - \frac{B(p^{2})-B(k^2)}{p^{2}-k^{2}}(p+k)_{\nu}.
\end{equation}

To perform the integration in 3-dimensional Euclidean space, we transcribe the Dyson-Schwinger equation by the following rules
\begin{align}
x^0 &\rightarrow -ix^0,\\
\mathbf{x}&\rightarrow \mathbf{x},\\
k^0 &\rightarrow -ik^0,\\
\mathbf{k}&\rightarrow \mathbf{k},
\end{align}
and
\begin{align}
\gamma^E_4 &= \gamma^0,\\
\gamma_i^E &= -i \gamma^{i},
\end{align}
so that
\begin{equation}
\{\gamma_{\mu},\gamma_{\nu}\} = 2 \delta_{\mu\nu}.
\end{equation}

Note that we also need to perform a reflection of $\mathbf{k}$ so that the free-fermion propagator has the form
\begin{equation}
S_0(p)= \frac{1}{i \gamma\cdot p + m}.
\end{equation}

Due to the spinor structure of fermion propagator, it admits the following general form
\begin{equation} \label{GeneralForm}
S^{-1}(p) = i p\!\!\!/ A(p^2) + B(p^2).
\end{equation}

Substitute Eq. \eqref{GeneralForm} into DSEs, we obtain the following set of couple integral equations
\begin{equation} \label{A1}
A(p^2) = 1 - \frac{1}{4p^2}\int \frac{d^3 k}{(2\pi)^2}Tr[i(\gamma P)\gamma_{\sigma}S(k)\Gamma_{\nu}D_{\sigma\nu}(q)],
\end{equation}
\begin{equation} \label{B1}
B(p^2) = \frac{1}{4}\int \frac{d^3(k)}{(2\pi)^3}Tr[\gamma_{\sigma}S(k)\Gamma_{\nu}D_{\sigma\nu}(q)].
\end{equation}

In Euclidean space, the photon propagator has the form
\begin{equation} \label{PiTensor1}
\Pi_{\mu\nu} = -\int \frac{d^3k}{(2\pi)^3}Tr[S(k)\gamma_{\mu}S(p)\Gamma_{\nu}(p,k)].
\end{equation}

A unit in which $e^2 = 1$ has been chosen,
it can be written
\begin{equation} \label{PiScalar}
\Pi(q)^{\mu\nu} = (q^2\delta^{\mu\nu} - q^{\mu}q^{\nu})\Pi(q^2),
\end{equation}
by the Ward-Takahashi identity
\begin{equation}
q_{\mu}\Pi^{\mu\nu}(q) = 0.
\end{equation}

The ultraviolet divergence in $\Pi(q^2)$ can be projected out by the following operator\cite{Projector}
\begin{equation}
\mathcal{P}_{\mu\nu} = \delta_{\mu\nu} - 3\frac{q_{\mu}}{q_{\nu}}{q^2}.
\end{equation}

Substitute the vertex into Eqs. \eqref{A1}, \eqref{B1} and \eqref{PiTensor1} and employ the trace techniques list in Appendix A, we arrive at
\begin{align}
A(p^{2}) %=& 1 - \frac{1}{4p^{2}}\int\frac{d^{3}k}{(2\pi)^{3}}\times Tr[i(\gamma p)\gamma_{\sigma}S(m,k)\Gamma_{\nu}(p,k)D_{\sigma\nu}(q)]\\
 =& 1 - \frac{1}{p^{2}}\int \frac{d^{3}k}{(2\pi)^{3}}  \left\{\frac{[A(p^{2})+A(k^{2})]A(k^{2})[(
p^2k^2-(p\cdot k)^2)/q^2-p\cdot k]}{[A^{2}(k^{2})k^{2}+B^{2}(k^{2})]q^{2}(1+\Pi(q^{2}))} \right. \nonumber\\
&+ \frac{[2(p^2k^2-(p\cdot k)^2)](B(p^2)-B(k^2))B(k^2)}{(A^{2}(k^{2})+B^{2}(k^{2}))q^{4}(1+\Pi(q^{2}))(p^{2}-k^{2})}    \nonumber\\
&+
 [2(k^2 + p^2)(k^2p^2 - (p \cdot k)^2)] \left.\frac{A(k^{2})[A(p^2)-A(k^2)]}{2(A^{2}(k^{2})k^{2}+B^{2}(k^{2}))(p^{2}-k^{2})q^{4}(1+\Pi(q^{2}))} \right \}, \label{A}\\
%  = &1 - \frac{1}{p^{2}}\int \frac{k^2d k}{(2\pi)^{2}} \frac{1}{(A^2(k^2)k^2+B^2(k^2))}\int sin\theta d\theta\left \{
%\frac{[A(p^{2})+A(k^{2})]A(k^{2})[(p^2k^2\,sin^2\theta)/q^2-pk\, cos\theta]}{q^2(1+\Pi(q^2))}\right.\\
%&\left. + [\frac{(B(p^2)-B(k^2))B(k^2)}{(p^{2}-k^{2})(1+\Pi(q^{2}))} + \frac{A(k^{2})[A(p^2)-A(k^2)]}{2(p^{2}-k^{2})(1+\Pi(q^{2}))}(k^2 + p^2)][2(k^2p^2\,sin^2\theta)/q^4]\right \} \\
B(p^{2})%& = \frac{1}{4} \int \frac{d^{3}k}{(2\pi)^{3}}Tr[\gamma_{\sigma}S(m,k)\Gamma_{\nu}(p,k)D_{\sigma\nu}(m,q)]  \\
% = &\int \frac{d^{3}k}{(2\pi)^{3}} \left \{ \frac{B(k^2)(A(p^2)+A(k^2))}{(A^2(k^2)k^2+B^2(k^2))q^2(1 + \Pi(q^2))} \right.\\
%&\left. + \frac{B(k^2)(A(p^2)-A(k^2))(2(p^2k^2-(p\cdot k)^2)/q^2)}{(A^2(k^2)k^2+B^2(k^2))q^2(p^2 - k^2)(1+\Pi(q^2))}- \frac{A(k^2)(B(p^2)-B(k^2))(2(p^2k^2-(p\cdot k)^2)/q^2)}{(A^2(k^2)k^2+B^2(k^2))q^2(p^2 - k^2)(1+\Pi(q^2))} \right \} \\
 =& \int \frac{d^{3}k}{(2\pi)^{3}}  \frac{1}{(A^2(k^2)k^2+B^2(k^2))(1+\Pi(q^2))}\left \{\frac{B(k^2)(A(p^2)+A(k^2))}{q^2} + \frac{B(k^2)A(p^2)-A(k^2)B(p^2)}{q^2(p^2 - k^2)}\right.\nonumber\\
 &\times\left.[2(p^2k^2-(p\cdot k)^2)/q^2)]\right\},   \label{B}\\
%=&\int \frac{k^2d k}{(2\pi)^{2}} \frac{1}{(A^2(k^2)k^2+B^2(k^2))} \times\\
%&\int \,sin\theta d\theta \frac{1}{(1+\Pi(q^2)}\left \{\frac{B(k^2)(A(p^2)+A(k^2))}{q^2} + \frac{B(k^2)A(p^2)-A(k^2)B(p^2)}{q^4(p^2 - k^2)}[2p^2k^2\,
%sin^2\theta]\right \}\\
\Pi(q^2) %=& -\frac{N_f}{2q^2}(\delta_{\sigma\nu}-3\frac{q_{\sigma}q_{\nu}}{q^2})\int \frac{d^3k}{(2\pi)^3}Tr[S(m,k)\gamma_{\sigma}S(m,q+k)\Gamma_{\nu}(m,p,k)] \\
 = &\frac{N_f}{q^2}\int \left\{ \frac{d^3k}{(2\pi)^3} \frac{A(p^2)A(k^2)[A(p^2)+A(k^2)]}{(A^2(k^2)k^2+B^2(k^2))(A^2(p^2)p^2+B^2(p^2))}[2k^2 - 4(k\cdot q)-6(k\cdot q)^2/q^2] \right. \nonumber\\
& + \frac{(A(p^2)-A(k^2))}{2(A^2(k^2)k^2+B(k^2))(A^2(p^2)p^2+B(p^2))(p^2 - k^2)}[4(B(k^2)B(p^2))(q^2 + 4k\cdot q + 6(k\cdot q)^2/q^2 - 2k^2)   \nonumber\\
& + 2A(k^2)A(p^2)(4k^4 - 4 k^2 k\cdot q - 10 (k \cdot q)^2 - 2 k \cdot q q^2 - 12 k^2 (k \cdot q)^2/q^2 - 12 (k \cdot q)^3/q^2)]    \nonumber\\
&+ 4[B(k^2)A(p^2)(k^2 - 3(k\cdot q)^2/q^2-3k\cdot q -q^2)
+ B(p^2)A(k^2)(k^2 - k\cdot q - 3(k\cdot q)^2/q^2)]        \nonumber\\
& \times \left. \frac{B(p^2)-B(k^2)}{(p^2-k^2)(A^2(k^2)k^2+B(k^2))(A^2(p^2)p^2+B(p^2))} \right\}. \label{C}\\ \nonumber
%= & \frac{N_f}{q^2} \int \frac{k^2d k}{(2\pi)^2} \frac{1}{(A^2(k^2)k^2+B(k^2))}\int\sin\theta d\theta\frac{1}{A^2(p^2)p^2+B(p^2)}\left\{ A(p^2)A(k^2)[A(p^2)+A(k^2)] \right.\\
%& \times[2k^2 - 4kq\, cos\theta-6(kq\, cos\theta)^2/q^2]+ \frac{(A(p^2)-A(k^2))}{(p^2 - k^2)}[2(B(k^2)B(p^2))(q^2 + 4kq\, cos\theta + 6(kq\, cos\theta)^2/q^2 - 2k^2)\\
%& + A(k^2)A(p^2)(4k^4-4k^2 kq\, cos\theta - 10 (kq\, cos\theta)^2 - 12k^2(kq\, cos\theta)^2/q^2 - 12(kq\, cos\theta)^3/q^2 - 2(kq\, cos\theta)q^2)]\\
%&+\left.\frac{B(p^2)-B(k^2)}{(p^2-k^2)}4[B(k^2)A(p^2)(k^2 - 3(kq \,cos\theta)^2/q^2-3kq \,cos\theta -q^2) +  B(p^2)A(k^2)(k^2 - k\, q\, cos\theta - 3 (k\, q\, cos\theta)^2/q^2)]\right\}
\end{align}
Our task is to find the numerical solution of these equations.
\section{Numerical method}

\subsection{Polynomial expansion}
As an improvement over the fixed-point integration, the polynomial expansion of the unknown function is adopted. This strategy is necessary if we want to calculate the values of function at arbitrary points (The fixed-point integration, as indicated by its name, only calculate values of functions at fixed point, and reduce the continuous values of functions to a discrete set of function values. In the calculation of complex vertex such as BC vertex, we will have to handle the singular integral kernel. We would encounter singular term proportional to $1/(p^2 - k^2)$. We may feel attempted to skip over the singular point of integration, however, this would cause the loss of accuracy and the iteration may never converge. On the other hand, if we expand the unknown function in terms of known polynomials-for example, the Chebychef polynomial-as long as the coefficients of expansion is known, we can calculate the value of the functions at any points. In this methods the coefficients will be calculate iteratively in Newton's method).
Chebychef polynomials have the following form
\begin{equation}
T_i(x) = cos(i\,acos(x)),
\end{equation}
where $x\in[-1, 1]$.

We also need to change the variable of integration to $ln(p^2)$ where $p$ is the momentum, to ensure sufficient sampling at singular points. Numerical cutoff is also needed so that the integration is done over the range $[\varepsilon, \Lambda]$. Therefore we arrive at the mapping
\begin{equation}
x = \frac{ln(p^2) - \frac{ln(\Lambda^2)+ln(\varepsilon^2)}{2}}{\frac{ln(\Lambda^2)-ln(\varepsilon^2)}{2}}.
\end{equation}

In general the unknown function can be expanded as
\begin{equation}
A(p^2) = \frac{a_0}{2} + \Sigma_{i=1}^{N} a_iT_{i}(x), \label{ExA}
\end{equation}
\begin{equation}
B(p^2) = \frac{b_0}{2} + \Sigma_{i=1}^{N} b_iT_{i}(x),  \label{ExB}
\end{equation}
\begin{equation}
\Pi(p^2) = \frac{c_0}{2} + \Sigma_{i=1}^{N} c_iT_{i}(x), \label{ExC}
\end{equation}
where $N$ is the number of terms in the approximation and $T_{i}$ is the $i\,th$ term of the Chebychef polynomial.
\subsection{Newton's method}
The Newton iteration
\begin{equation}
x_{n+1} = x_n - F^{\prime}(x_{n})^{-1}F(x_{n}),
\end{equation}
exhibits q-quadratical convergence \cite{QQuad}. Therefore it was chosen as the basis of the algorithm. However, we would have to deal with the addition complication of the derivatives of the unknown functions with respect to the coefficients of the expansions. The integration in Eqs. \eqref{A}, \eqref{B} and \eqref{C} can expressed in the spherical coordinate in 3-dimensional Euclidean space
\begin{align}
A(p^2)= &1 - \frac{1}{p^{2}}\int \frac{k^2d k}{(2\pi)^{2}} \frac{1}{(A^2(k^2)k^2+B^2(k^2))}\int sin\theta d\theta\left \{
\frac{[A(p^{2})+A(k^{2})]A(k^{2})[(p^2k^2\,sin^2\theta)/q^2-pk\, cos\theta]}{q^2(1+\Pi(q^2))}\right.\nonumber\\
&\left. + [\frac{(B(p^2)-B(k^2))B(k^2)}{(p^{2}-k^{2})(1+\Pi(q^{2}))} + \frac{A(k^{2})[A(p^2)-A(k^2)]}{2(p^{2}-k^{2})(1+\Pi(q^{2}))}(k^2 + p^2)][2(k^2p^2\,sin^2\theta)/q^4]\right \}, \label{A2}\\
B(p^2)=&\int \frac{k^2d k}{(2\pi)^{2}} \frac{1}{(A^2(k^2)k^2+B^2(k^2))} \times  \nonumber\\
&\int \,sin\theta d\theta \frac{1}{(1+\Pi(q^2)}\left \{\frac{B(k^2)(A(p^2)+A(k^2))}{q^2} + \frac{B(k^2)A(p^2)-A(k^2)B(p^2)}{q^4(p^2 - k^2)}[2p^2k^2\,
sin^2\theta]\right \}, \label{B2}\\
\Pi(p^2)= & \frac{N_f}{q^2} \int \frac{k^2d k}{(2\pi)^2} \frac{1}{(A^2(k^2)k^2+B(k^2))}\int\sin\theta d\theta\frac{1}{A^2(p^2)p^2+B(p^2)}\left\{ A(p^2)A(k^2)[A(p^2)+A(k^2)] \right. \nonumber\\
& \times[2k^2 - 4kq\, cos\theta-6(kq\, cos\theta)^2/q^2]+ \frac{(A(p^2)-A(k^2))}{(p^2 - k^2)}[2(B(k^2)B(p^2))(q^2 + 4kq\, cos\theta + 6(kq\, cos\theta)^2/q^2 - 2k^2)  \nonumber\\
& + A(k^2)A(p^2)(4k^4-4k^2 kq\, cos\theta - 10 (kq\, cos\theta)^2 - 12k^2(kq\, cos\theta)^2/q^2 - 12(kq\, cos\theta)^3/q^2 - 2(kq\, cos\theta)q^2)]  \nonumber\\
&+\left.\frac{B(p^2)-B(k^2)}{(p^2-k^2)}4[B(k^2)A(p^2)(k^2 - 3(kq \,cos\theta)^2/q^2-3kq \,cos\theta -q^2)\right.\nonumber\\
& +  \left.B(p^2)A(k^2)(k^2 - k\, q\, cos\theta - 3 (k\, q\, cos\theta)^2/q^2)]\right\}, \label{C2}
\end{align}

Substitute the expansions \eqref{ExA}, \eqref{ExC} and \eqref{ExB}, we obtain the following nonlinear system of integral equation
\begin{align}
F_A(a_i, b_i, c_i) = 0,\\
F_B(a_i, b_i, c_i) = 0,\\
F_{\Pi}(a_i, b_i, c_i) = 0,
\end{align}
where
\begin{align}
F_A(a_i, b_i, c_i) =& A(p^2) - 1 + \frac{1}{p^{2}}\int \frac{k^2d k}{(2\pi)^{2}} \frac{1}{(A^2(k^2)k^2+B^2(k^2))}\int sin\theta d\theta\left \{
\frac{[A(p^{2})+A(k^{2})]A(k^{2})[(p^2k^2\,sin^2\theta)/q^2-pk\, cos\theta]}{q^2(1+\Pi(q^2))}\right.\nonumber\\
&\left. + [\frac{(B(p^2)-B(k^2))B(k^2)}{(p^{2}-k^{2})(1+\Pi(q^{2}))} + \frac{A(k^{2})[A(p^2)-A(k^2)]}{2(p^{2}-k^{2})(1+\Pi(q^{2}))}(k^2 + p^2)][2(k^2p^2\,sin^2\theta)/q^4]\right \} \label{A3},\\
F_B(a_i, b_i, c_i) =& B(p^2)-\int \frac{k^2d k}{(2\pi)^{2}} \frac{1}{(A^2(k^2)k^2+B^2(k^2))} \times  \nonumber\\
&\int \,sin\theta d\theta \frac{1}{(1+\Pi(q^2)}\left \{\frac{B(k^2)(A(p^2)+A(k^2))}{q^2} + \frac{B(k^2)A(p^2)-A(k^2)B(p^2)}{q^4(p^2 - k^2)}[2p^2k^2\,
sin^2\theta]\right \}, \label{B3}\\
F_{\Pi}(a_i, b_i, c_i) = &\Pi(p^2) -  \frac{N_f}{q^2} \int \frac{k^2d k}{(2\pi)^2} \frac{1}{(A^2(k^2)k^2+B(k^2))}\int\sin\theta d\theta\frac{1}{A^2(p^2)p^2+B(p^2)}\left\{ A(p^2)A(k^2)[A(p^2)+A(k^2)] \right. \nonumber\\
& \times[2k^2 - 4kq\, cos\theta-6(kq\, cos\theta)^2/q^2]+ \frac{(A(p^2)-A(k^2))}{(p^2 - k^2)}\nonumber\\
&\times[2(B(k^2)B(p^2))(q^2 + 4kq\, cos\theta + 6(kq\, cos\theta)^2/q^2 - 2k^2)  \nonumber\\
& + A(k^2)A(p^2)(4k^4-4k^2 kq\, cos\theta - 10 (kq\, cos\theta)^2 - 12k^2(kq\, cos\theta)^2/q^2 - 12(kq\, cos\theta)^3/q^2 - 2(kq\, cos\theta)q^2)]  \nonumber\\
&+\frac{B(p^2)-B(k^2)}{(p^2-k^2)}4[B(k^2)A(p^2)(k^2 - 3(kq \,cos\theta)^2/q^2-3kq \,cos\theta -q^2) \nonumber\\
&+\left.B(p^2)A(k^2)(k^2 - k\, q\, cos\theta - 3 (k\, q\, cos\theta)^2/q^2)]\right\}, \label{C3}
\end{align}
and
\begin{equation*}
A(p^2) = \frac{a_0}{2} + \Sigma_{i=1}^{N} a_iT_{i}(x) \label{ExA2},
\end{equation*}
\begin{equation*}
B(p^2) = \frac{b_0}{2} + \Sigma_{i=1}^{N} b_iT_{i}(x)  \label{ExB2},
\end{equation*}
\begin{equation*}
\Pi(p^2) = \frac{c_0}{2} + \Sigma_{i=1}^{N} c_iT_{i}(x) \label{ExC2}.
\end{equation*}

We next calculate the derivatives of $F_{A}(a_i, b_i, c_i), F_{B}(a_i, b_i, c_i)$ and $F_{\Pi}(a_i, b_i, c_i)$ with respect to $a_i, b_i$ and $c_i$, respectively. To clarify the calculation, the following notations are adopted
\begin{align}
T_1 &= (p^2k^2\,sin^2\theta)/q^2-pk\, cos\theta, \\
T_2 &= 2(k^2p^2\,sin^2\theta)/q^2,   \\
%T_3 &= p^2k^2\,sin^2\theta \\
T_3 &= 2k^2 - 4kq\, cos\theta-6(kq\, cos\theta)^2/q^2, \\
T_4 &= q^2 + 4kq\, cos\theta + 6(kq\, cos\theta)^2/q^2 - 2k^2,\\
T_5 &= 4k^4-4k^2 kq\, cos\theta - 10 (kq\, cos\theta)^2 - 12k^2(kq\, cos\theta)^2/q^2 - 12(kq\, cos\theta)^3/q^2 - 2(kq\, cos\theta)q^2,        \\
T_6 &= k^2 - 3(kq \,cos\theta)^2/q^2-3kq \,cos\theta -q^2,  \\
T_7 &= k^2 - k\, q\, cos\theta - 3 (k\, q\, cos\theta)^2/q^2,
\end{align}
then we have
\begin{align}
\frac{\partial{F_{A}}}{\partial{a_i}} =& T_i(x) + \int\frac{k^2sin\theta\, dk\,d\theta}{(2\pi)^2}\left\{\frac{A(k^2)}{p^2q^2(1+\Pi(q^2))(A^2(k^2)k^2+B^2(k^2))} T_1\,T_i(x)\right. \nonumber\\
& + \frac{A(k^2)(p^2+k^2)}{2(p^2-k^2)p^2q^2  (1+\Pi(q^2))(A^2(k^2)k^2+B^2(k^2))}T_2\,T_i(x)\nonumber \\
& -\frac{2A^2(k^2)(A(k^2)+A(p^2))k^2}{p^2q^2 (1+\Pi(q^2))(A^2(k^2)k^2+B^2(k^2))^2}T_1T_i(y)
-\frac{2A(k^2)B(k^2)(B(p^2)-B(k^2))k^2}{(p^2-k^2)p^2q^2(1+\Pi(q^2))(A^2(k^2)k^2+B^2(k^2))^2}T_2T_i(y) \nonumber \\
& -\frac{A^2(k^2)(A(p^2)-A(k^2))(p^2 + k^2)k^2}{(p^2-k^2)p^2q^2(1+\Pi(q^2))(A^2(k^2)k^2+B^2(k^2))^2}T_2T_i(y)
+\frac{A(k^2)}{p^2q^2(1+\Pi(q^2))(A^2(k^2)k^2+B^2(k^2))}T_1T_j(y)      \nonumber   \\
&+ \frac{(A(k^2)+A(p^2))}{p^2q^2(1+\Pi(q^2))(A^2(k^2)k^2+B^2(k^2))}T_1T_i(y)
-\frac{A(k^2)(p^2+k^2)}{2(p^2-k^2)p^2q^2(1+\Pi(q^2))(A^2(k^2)k^2+B^2(k^2))}T_2T_i(y)  \nonumber \\
&+ \left.\frac{A(p^2)-A(k^2)}{2(p^2-k^2)p^2q^2(1+\Pi(q^2))(A^2(k^2)k^2+B^2(k^2))}T_2T_i(y) \right\},
\end{align}
\begin{align}
\frac{\partial{F_{A}}}{\partial{b_i}} =& \int\frac{k^2sin\theta\, dk\,d\theta}{(2\pi)^2}\left\{\frac{B(k^2)}{(p^2-k^2)p^2q^2(1+\Pi(q^2))(A^2(k^2)k^2+B^2(k^2))}T_2T_i(x)\right.\nonumber\\
&-\left.\frac{2A(k^2)(A(k^2)+A(p^2))B(k^2)}{p^2q^2(1+\Pi(q^2))(A^2(k^2)k^2+B^2(k^2))^2}\right. \nonumber\\
&\times T_1T_i(y) -\frac{2B(k^2)(B(p^2)-B(k^2))}{(p^2-k^2)p^2q^2(1+\Pi(q^2))(A^2(k^2)k^2+B^2(k^2))^2}T_2T_i(y) \nonumber \\
&-\frac{A(k^2)B(k^2)(A(p^2)-A(k^2))(p^2+k^2)}{(p^2-k^2)p^2q^2(1+\Pi(q^2))(A^2(k^2)k^2+B^2(k^2))^2}T_2T_i(y)\nonumber\\
&-\frac{B(k^2)}{(p^2-k^2)p^2q^2(1+\Pi(q^2))(A^2(k^2)k^2+B^2(k^2))}T_2T_i(y)\nonumber\\
&+\left.\frac{B(p^2)-B(k^2)}{(p^2-k^2)p^2q^2(1+\Pi(q^2))(A^2(k^2)k^2+B^2(k^2))}T_2T_i(y) \right\},
\end{align}
\begin{align}
\frac{\partial{F_A}}{\partial{c_i}} =& \int\frac{k^2sin\theta\, dk\,d\theta}{(2\pi)^2}\left\{-\frac{A(k^2)(A(k^2)+A(p^2))}{p^2q^2(1+\Pi(q^2))^2(A^2(k^2)k^2+B^2(k^2))}T_1T_i(z) \right.\nonumber\\
&- \left.\frac{B(k^2)(B(p^2)-B(k^2))}{(p^2-k^2)p^2q^2(1+\Pi(q^2))^2(A^2(k^2)k^2+B^2(k^2))}T_2T_i(z) \right.\nonumber\\
&- \left.\frac{A(k^2)(A(p^2)-A(k^2))(p^2 + k^2)}{2(p^2-k^2)p^2q^2(1+\Pi(q^2))^2(A^2(k^2)k^2+B^2(k^2))}T_2T_i(z)\right\},
\end{align}
\begin{align}
\frac{\partial{F_B}}{\partial{a_i}} =&\int\frac{k^2sin\theta\, dk\,d\theta}{(2\pi)^2}\left\{ -\frac{B(k^2)}{q^2(1+\Pi(q^2))(A^2(k^2)k^2+B^2(k^2))}T_i(x)\right.\nonumber\\
&- \frac{B(k^2)}{(p^2-k^2)q^2(1+\Pi(q^2))(A^2(k^2)k^2+B^2(k^2))}T_2T_i(x) \nonumber\\
&+ \frac{2A(k^2)(A(k^2)+A(p^2))B(k^2)k^2}{q^2(1+\Pi(q^2))(A^2(k^2)k^2+B^2(k^2))^2}T_i(y)
+ \frac{2A(k^2)(A(p^2)B(k^2)-A(k^2)B(p^2))}{(p^2-k^2)q^2(1+\Pi(q^2))(A^2(k^2)k^2+B^2(k^2))^2}T_2T_i(y)  \nonumber\\
&\left.-\frac{B(k^2)}{q^2(1+\Pi(q^2))(A^2(k^2)k^2+B^2(k^2))}T_i(y)
+ \frac{B(p^2)}{(p^2-k^2)q^2(1+\Pi(q^2))(A^2(k^2)k^2+B^2(k^2))}T_2T_i(y)\right\},
\end{align}
\begin{align}
\frac{\partial{F_B}}{\partial{b_i}} =& T_i(x) + \int\frac{k^2sin\theta\, dk\,d\theta}{(2\pi)^2}\left\{\frac{A(k^2)}{(p^2-k^2)(1+\Pi(q^2))q^2(A^2(k^2)k^2+B^2(k^2))}T_2T_i(x)\right.  \nonumber\\
&+\frac{2(A(k^2)+A(k^2))B(k^2)}{q^2(1+\Pi(q^2))(A^2(k^2)k^2+B^2(k^2))^2}T_2
+ \frac{2B(k^2)(A(p^2)B(k^2)-A(k^2)B(p^2))}{(p^2-k^2)q^2(1+\Pi(q^2))(A^2(k^2)k^2+B^2(k^2))^2}T_2T_i(y) \nonumber\\
& -\left. \frac{A(k^2)+A(p^2)}{q^2(1+\Pi(q^2))(A^2(k^2)k^2+B^2(k^2))}T_2
- \frac{A(p^2)}{(p^2-k^2)q^2(1+\Pi(q^2))(A^2(k^2)k^2+B^2(k^2))}T_2T_i(y) \right\},
\end{align}
\begin{align}
\frac{\partial{F_B}}{\partial{c_i}} =&\int\frac{k^2sin\theta\, dk\,d\theta}{(2\pi)^2}\left\{ \frac{(A(k^2)+A(p^2))B(k^2)}{q^2(1+\Pi(q^2))^2(A^2(k^2)k^2+B^2(k^2))}T_i(z)
\right.\nonumber\\
&+\left.\frac{(A(p^2)B(k^2)-A(k^2)B(p^2))}{(p^2-k^2)q^2(1+\Pi(q^2))^2(A^2(k^2)k^2+B^2(k^2))}T_2T_i(z)\right\}, \\
\frac{\partial{F_{\Pi}}}{\partial{a_i}} =& \int\frac{k^2sin\theta\, dk\,d\theta}{(2\pi)^2}\left\{\frac{2A(k^2)A(p^2)(A(k^2)k^2+A(p^2))p^2}{q^2(A^2(p^2)p^2+B(p^2))^2(A^2(k^2)k^2+B^2(k^2))}T_3T_i(x)\right.\nonumber\\
&-\frac{A(k^2)A(p^2)}{q^2(A^2(p^2)p^2+B(p^2))(A^2(k^2)k^2+B^2(k^2))}T_3T_i(x) \nonumber\\
&-\frac{A(k^2)(A(k^2)+A(p^2))}{q^2(A^2(p^2)p^2+B(p^2))(A^2(k^2)k^2+B^2(k^2))}T_3T_i(x)\nonumber\\
&+\frac{4A(p^2)(A(p^2)-A(k^2))B(k^2)B(p^2)p^2}{(p^2-k^2)q^2(A^2(p^2)p^2+B(p^2))^2(A^2(k^2)k^2+B^2(k^2))}T_4T_i(x)\nonumber \\
&-\frac{2B(k^2)B(p^2)}{(p^2-k^2)q^2(A^2(p^2)p^2+B(p^2))(A^2(k^2)k^2+B^2(k^2))}T_4T_i(x)\nonumber\\
&+\frac{2A(k^2)A^2(p^2)(A(p^2)-A(k^2))p^2}{(p^2-k^2)q^2(A^2(p^2)p^2+B(p^2))^2(A^2(k^2)k^2+B^2(k^2))}T_5T_i(x)\nonumber\\
&-\frac{A(k^2)A(p^2)}{(p^2-k^2)q^2(A^2(p^2)p^2+B(p^2))(A^2(k^2)k^2+B^2(k^2))}T_5T_i(x)\nonumber\\
&-\frac{A(k^2)(A(p^2)-A(k^2))}{(p^2-k^2)q^2(A^2(p^2)p^2+B(p^2))(A^2(k^2)k^2+B^2(k^2))}T_5T_i(x)  \nonumber\\
%\end{align}
%\begin{align}
&+\frac{8A^2(p^2)B(k^2)(B(p^2)-B(k^2))p^2}{(p^2-k^2)q^2(A^2(p^2)p^2+B(p^2))^2(A^2(k^2)k^2+B^2(k^2))}T_6T_i(x)\nonumber\\
&-\frac{4B(k^2)(B(p^2)-B(k^2))}{(p^2-k^2)q^2(A^2(p^2)p^2+B(p^2))(A^2(k^2)k^2+B^2(k^2))}T_6T_i(x) \nonumber\\
&+\frac{8A(k^2)A(p^2)B(p^2)(B(p^2)-B(k^2))p^2}{(p^2-k^2)q^2(A^2(p^2)p^2+B(p^2))^2(A^2(k^2)k^2+B^2(k^2))}T_7T_i(x)\nonumber\\
&+ \frac{2A^2(k^2)A(p^2)(A(k^2)+A(p^2))k^2}{q^2(A^2(p^2)p^2+B(p^2))(A^2(k^2)k^2+B^2(k^2))^2}T_3T_i(y) \nonumber \\
&+\frac{4A(k^2)(A(p^2)-A(k^2))B(k^2)B(p^2)k^2}{(p^2-k^2)q^2(A^2(p^2)p^2+B(p^2))(A^2(k^2)k^2+B^2(k^2))^2}T_4T_i(y)\nonumber\\
&+\frac{2A^2(k^2)A(p^2)(A(p^2)-A(k^2))k^2}{(p^2-k^2)q^2(A^2(p^2)p^2+B(p^2))(A^2(k^2)k^2+B^2(k^2))^2}T_5T_i(y) \nonumber\\
&+ \frac{8A(k^2)A(p^2)B(k^2)(B(p^2)-B(k^2))k^2}{(p^2-k^2)q^2(A^2(p^2)p^2+B(p^2))(A^2(k^2)k^2+B^2(k^2))^2}T_6T_i(y)\nonumber\\
&-\frac{A(k^2)A(p^2)}{q^2(A^2(p^2)p^2+B(p^2))(A^2(k^2)k^2+B^2(k^2))}T_3T_i(y) \nonumber \\
&-\frac{A(p^2)(A(k^2)+A(p^2))}{q^2(A^2(p^2)p^2+B(p^2))(A^2(k^2)k^2+B^2(k^2))}T_3T_i(y) \nonumber\\
&+ \frac{2B(k^2)B(p^2)}{(p^2-k^2)q^2(A^2(p^2)p^2+B(p^2))(A^2(k^2)k^2+B^2(k^2))}T_4T_i(y) \nonumber\\
&+ \frac{A(k^2)A(p^2)}{(p^2-k^2)q^2(A^2(p^2)p^2+B(p^2))(A^2(k^2)k^2+B^2(k^2))}   T_5T_i(y)\nonumber\\
&- \frac{A(p^2)(A(p^2)-A(k^2))}{(p^2-k^2)q^2(A^2(p^2)p^2+B(p^2))(A^2(k^2)k^2+B^2(k^2))}T_5T_i(y)\nonumber \\
&- \left.\frac{4B(p^2)(B(p^2)-B(k^2))}{(p^2-k^2)q^2(A^2(p^2)p^2+B(p^2))(A^2(k^2)k^2+B^2(k^2))}T_7T_i(y)  \right\},
\end{align}
\begin{align}
\frac{\partial{F_{\Pi}}}{\partial{b_i}} =& \int\frac{k^2sin\theta\, dk\,d\theta}{(2\pi)^2}\left\{\frac{2A(k^2)A(p^2)(A(k^2)+A(p^2))B(p^2)}{q^2(A^2(p^2)p^2+B(p^2))^2(A^2(k^2)k^2+B^2(k^2))}T_3T_i(x)\right.\nonumber\\
&+\frac{4(A(p^2)-A(k^2))B(k^2)B^2(p^2)}{(p^2-k^2)q^2(A^2(p^2)p^2+B(p^2))^2(A^2(k^2)k^2+B^2(k^2))}T_4T_i(x) \nonumber \\
&-\frac{2(A(p^2)-A(k^2))B(k^2)}{(p^2-k^2)q^2(A^2(p^2)p^2+B(p^2))(A^2(k^2)k^2+B^2(k^2))}T_4T_i(x)\nonumber\\
&+ \frac{2A(k^2)A(p^2)(A(p^2)-A(k^2))B(p^2)}{(p^2-k^2)q^2(A^2(p^2)p^2+B(p^2))^2(A^2(k^2)k^2+B^2(k^2))}T_5T_i(x)\nonumber \\
&+\frac{8A(p^2)B(k^2)B(p^2)(B(p^2)-B(k^2))}{(p^2-k^2)q^2(A^2(p^2)p^2+B(p^2))^2(A^2(k^2)k^2+B^2(k^2))}T_6T_i(x)\nonumber \\
&- \frac{4A(p^2)B(k^2)}{(p^2-k^2)q^2(A^2(p^2)p^2+B(p^2))(A^2(k^2)k^2+B^2(k^2))}   T_6T_i(x) \nonumber \\
&+ \frac{8A(k^2)B(p^2)(B(p^2)-B(k^2))}{(p^2-k^2)q^2(A^2(p^2)p^2+B(p^2))^2(A^2(k^2)k^2+B^2(k^2))}T_7T_i(x)\nonumber\\
&- \frac{4A(k^2)B(p^2)}{(p^2-k^2)q^2(A^2(p^2)p^2+B(p^2))(A^2(k^2)k^2+B^2(k^2))}T_7T_i(x) \nonumber \\
& -\frac{4A(k^2)(B(p^2)-B(k^2))}{(p^2-k^2)q^2(A^2(p^2)p^2+B(p^2))(A^2(k^2)k^2+B^2(k^2))}T_7T_i(x)\nonumber\\
%\end{align}
%\begin{align}
&+ \frac{2A(k^2)A(p^2)(A(k^2)+A(p^2))B(k^2)}{q^2(A^2(p^2)p^2+B(p^2))(A^2(k^2)k^2+B^2(k^2))^2}T_3T_i(y) \nonumber\\
&+ \frac{4(A(p^2)-A(k^2))B^2(k^2)B(p^2)}{(p^2-k^2)q^2(A^2(p^2)p^2+B(p^2))(A^2(k^2)k^2+B^2(k^2))^2}T_4T_i(y)\nonumber\\
&+ \frac{2A(k^2)A(p^2)(A(p^2)-A(k^2))B(k^2)}{(p^2-k^2)q^2(A^2(p^2)p^2+B(p^2))(A^2(k^2)k^2+B^2(k^2))^2}T_5T_i(y) \nonumber\\
&+ \frac{8A(p^2)B^2(k^2)(B(p^2)-B(k^2))}{(p^2-k^2)q^2(A^2(p^2)p^2+B(p^2))(A^2(k^2)k^2+B^2(k^2))^2}T_6T_i(y)\nonumber\\
&- \frac{2(A(p^2)-A(k^2))B(p^2)}{(p^2-k^2)q^2(A^2(p^2)p^2+B(p^2))(A^2(k^2)k^2+B^2(k^2))}T_4T_i(y)   \nonumber\\
&+ \frac{4A(p^2)B(k^2)}{(p^2-k^2)q^2(A^2(p^2)p^2+B(p^2))(A^2(k^2)k^2+B^2(k^2))}T_6T_i(y)  \nonumber\\
&- \frac{4A(p^2)(B(p^2)-B(k^2))}{(p^2-k^2)q^2(A^2(p^2)p^2+B(p^2))(A^2(k^2)k^2+B^2(k^2))}T_6T_i(y) \nonumber\\
&+ \frac{4A(k^2)B(p^2)(B(p^2)-B(k^2))(2B(k^2)+2B(k^2)k^2)}{(p^2-k^2)q^2(A^2(p^2)p^2+B(p^2))(A^2(k^2)k^2+B^2(k^2))^2}T_7T_i(y)  \nonumber\\
&+ \left.\frac{4A(k^2)B(p^2)}{(p^2-k^2)q^2(A^2(p^2)p^2+B(p^2))(A^2(k^2)k^2+B^2(k^2))} \right\}T_7T_i(y),\\
\frac{\partial{F_{\Pi}}}{\partial{c_i}} =& T_i(z),
\end{align}
where
\begin{equation}
x = \frac{ln(p^2) - \frac{ln(\Lambda^2)+ln(\varepsilon^2)}{2}}{\frac{ln(\Lambda^2)-ln(\varepsilon^2)}{2}}, y = \frac{ln(k^2) - \frac{ln(\Lambda^2)+ln(\varepsilon^2)}{2}}{\frac{ln(\Lambda^2)-ln(\varepsilon^2)}{2}}, z = \frac{ln(q^2) - \frac{ln(\Lambda^2)+ln(\varepsilon^2)}{2}}{\frac{ln(\Lambda^2)-ln(\varepsilon^2)}{2}}.
\end{equation}

To avoid the singular points, in integration is split as
\begin{equation}
\int_{\varepsilon}^{\Lambda} = \int_{\varepsilon}^{p} + \int_{p}^{\Lambda},
\end{equation}
where $p$ is the external momentum. In the numerical solution, n-point Gaussian quadrature rule over the internal momentum $k$ is used. The number of Gaussian points on two ranges $[\varepsilon, p]$ and $[p,\Lambda]$ are set to $N_1 = 50$ and $N_2 = 50$. In addition, the number of integration over the angle $\theta$ is set to $N_{\theta}=40$
\subsection{Global convergence}
The newton's method is sensitive to the initial values of iteration, in this article a method with global convergence is presented.
The iteration of Newton's method involves the iteration\cite{Thesis}
\begin{equation}
J(x_n)\Delta_{n+1} = F(x_n),
\end{equation}
where
\begin{equation}
x = \left(\begin{array}{c}a_i\\b_i\\c_i\end{array}\right),\ \ \ \  F(x) = \left(\begin{array}{c}F_A(a_i,b_i,c_i)\\F_B(a_i,b_i,c_i)\\F_{\Pi}(a_i,b_i,c_i)\end{array}\right), \ \ \ \ J(x)=\frac{\partial{F_i}(x)}{\partial{x_j}}, \ \ \ \ \Delta_{n+1}= x_n - x_{n+1},
\end{equation}
with $n$ indicating the number of iteration.
The global converging method modifies the iteration $x_{n+1} = x_n - \Delta_{n+1}$ by setting\cite{QQuad}
\begin{equation}
x_{n+1} = x_n - \lambda\Delta_{n+1},
\end{equation}
where $\lambda\in(0,1]$. This modification gives us the flexibility in adjusting the step length of iteration.

The $\Delta_{n}$ and $F(x_n)$ should approach to $0$ in the limit $n\rightarrow \infty$, when $x$ is assumed to converge to the exact value. Since $F(x_n)$ should approach $0$ after sufficient amount of steps of iteration, we expect that our method should reduce the absolute value $||F(x_n)||_2$ after every iteration.This can be achieved by carefully adjusting the step length of iteration through choosing appropriate $\lambda$. We set the condition for convergence to be $||F(x_n)||_2 < \tau_a$ where $\tau_{a}$ is the predetermined absolute tolerance. We also need a relative tolerance $\tau_r$ as the condition for convergence. Thus the final $F(x_{n})$ should satisfy
\begin{equation}
||F(x_n)||_2 \leq \tau_r||F_0(x_n)||_2 + \tau_a,
\end{equation}
where $F_0(x_n)$ is the exact value.

To avoid possible stagnancy in the iteration, we need two parameter $\sigma_1$ and $\sigma_2$, with $0<\sigma_1<\sigma_2<1$, then we have the following safeguard against possible stagnancy in the iteration. $\lambda$ in the current trial step should satisfy\cite{QQuad}
\begin{equation}
\lambda = \left \{
\begin{array}{cc}
\sigma_1\lambda_1, &  \lambda < \sigma_1\lambda_1 \\
\lambda,          &  \sigma_1\lambda_1 < \lambda <\sigma_2\lambda_1\\
\sigma_2\lambda_1, &  \sigma_2\lambda_1 < \lambda
\end{array} \right.,
\end{equation}
where $\lambda_1$ is the value in the previous trial step.
\subsection{Three-point parabolic model}
In the algorithm, we first set $\lambda = 1$ to iterate. In a trial step, if the $\lambda = 1$ fail to reduce the absolute value $||\Delta_n||_2$, then, we set $\lambda = \sigma_1 \lambda$. When we have two successive trials that fail to reduce the value of $||\Delta_n||_2$, the next $\lambda$ can be determined by
the three-point parabolic model
\begin{equation}
\lambda = - \frac{(\lambda_1  (F2 - F0) / \lambda_2 - \lambda_2  (F1 - F0) / \lambda_1) / (\lambda_1 - \lambda_2)}{2((F1 - F0) / \lambda_1 - (F2 - F0) / \lambda_2) / (\lambda_1 - \lambda_2)},
\end{equation}
where
\begin{equation}
F0 = F(x_n), \ \ \ \  F1 = F(x_n - \lambda_1 \Delta_{n+1}), \ \ \ \  F2 = F(x_n - \lambda_2 \Delta_{n+1}).
\end{equation}
\section{Numerical results}
For comparison, the unknown functions $A(p^2), B(p^2), \Pi(p^2)$ with quenched approximation, one-loop contribution, and the minimal BC vertex (namely the first term in the BC vertex, which introduces the wavefunction renormalization dependence in the vertex when compared to bare approximation) are also calculated.
\subsection{Quenched approximation}
In the quenched approximation, we have
\begin{equation}
\Pi(q^2)=0,
\end{equation}
i.e. we neglect the fermion loop contribution to the vacuum polarization and adopt BC vertex in the calculation. Under this approximation, we calculate the $A(p^2), B(p^2), \Pi(p^2)$ using the methods stated above. Then we have
\begin{figure}[h]
\includegraphics[width = 100mm]{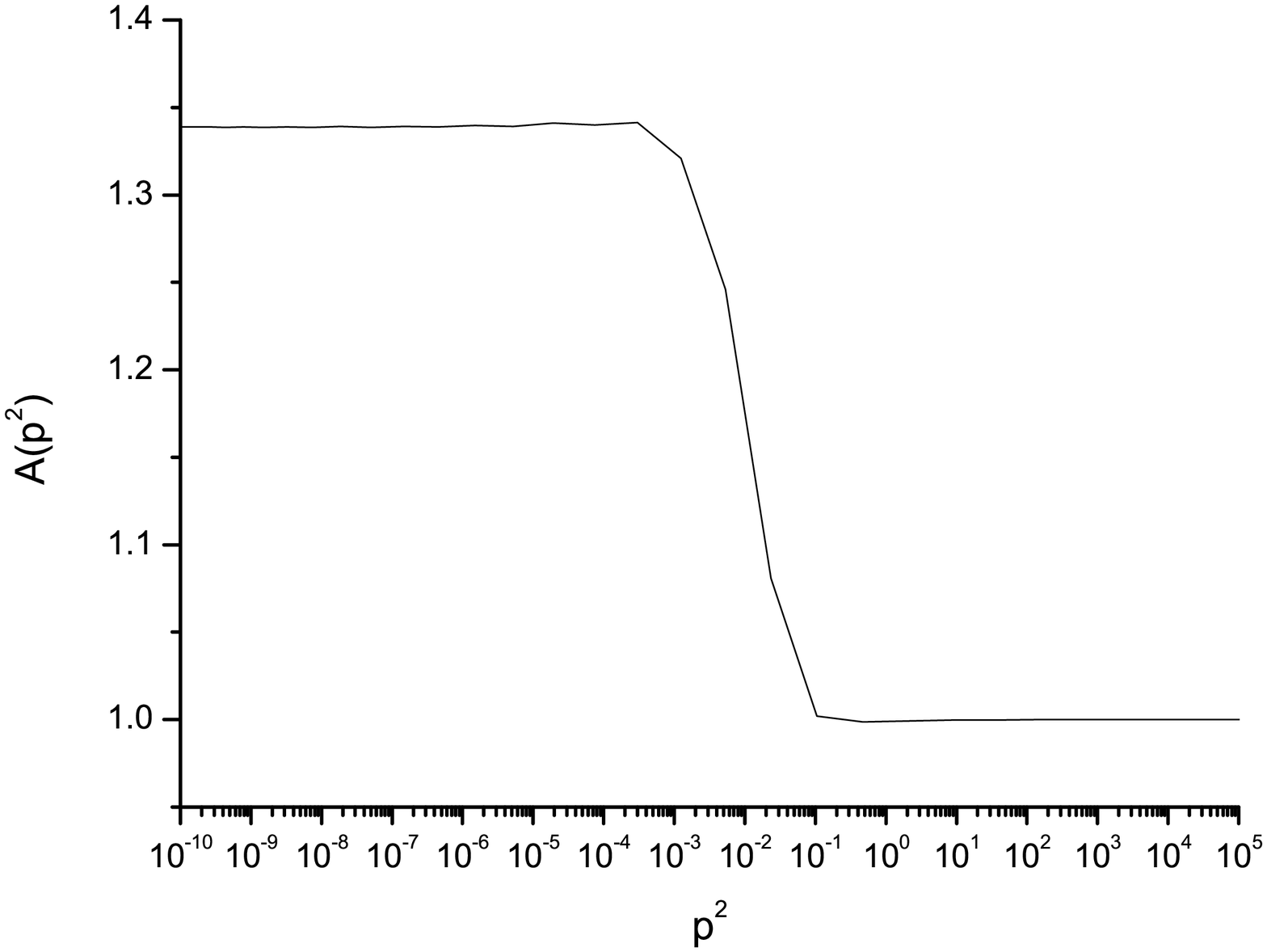}
\includegraphics[width = 100mm]{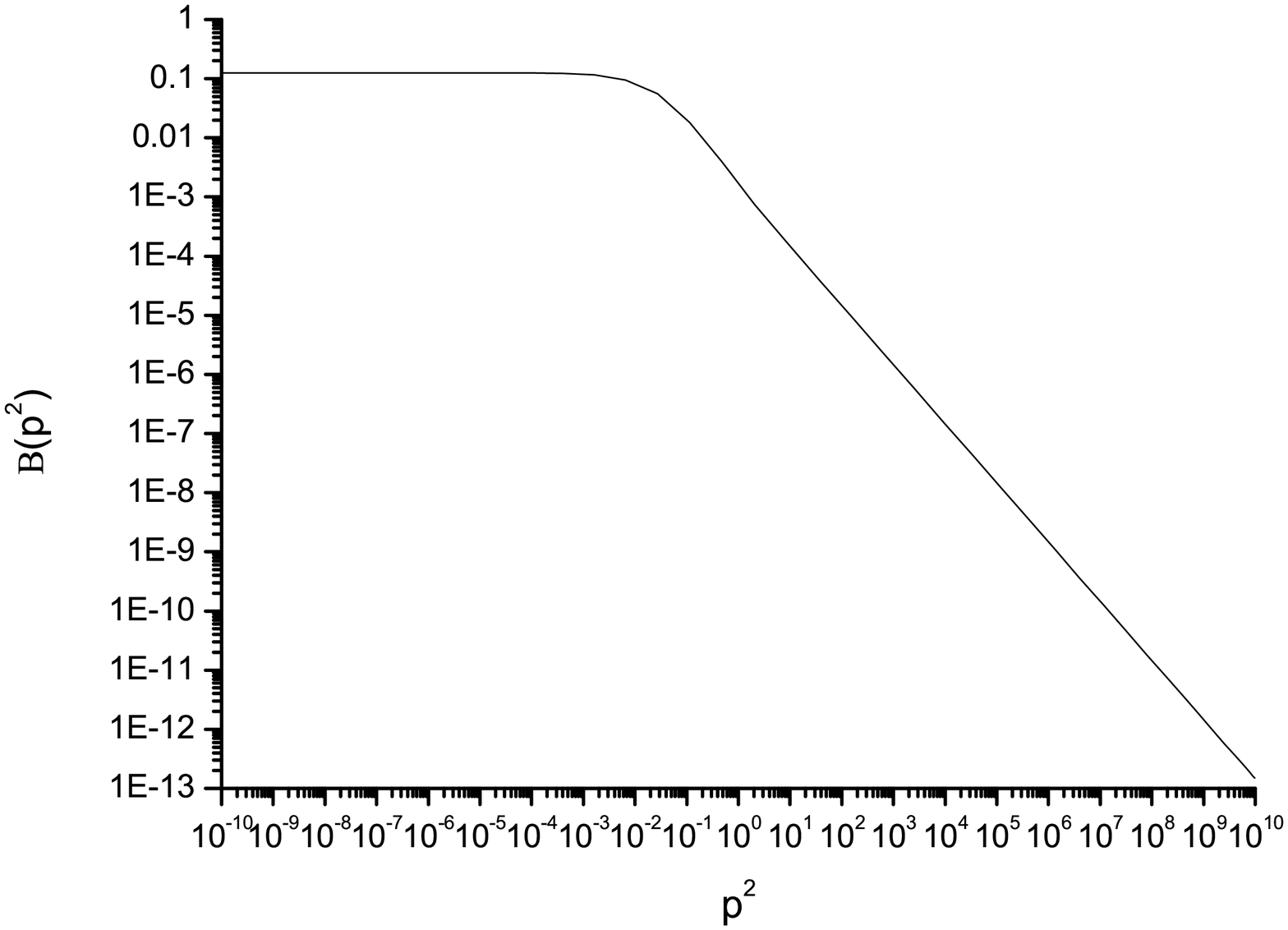}
\caption{The behavior of $A(p^2)$ and $B(p^2)$ with $N_f$ = 1 and $\Pi(p^2) \equiv 0$.}
\end{figure}
\subsection{One-loop approximation}
In the one-loop, we only consider the one-loop contribution of fermion to the vacuum polarization
\begin{equation}
\Pi_{\mu\nu}(q) = -\int \frac{d^3k}{(2\pi)^3}Tr[\frac{1}{i\gamma\cdot k}\gamma_{\mu}\frac{1}{i\gamma\cdot p}\gamma_{\nu}].
\end{equation}

Also we can factor out the tensor structure of the vacuum polarization
\begin{equation}
\Pi_{\mu\nu}(q) = (\delta_{\mu\nu}q^2 - q_{\mu}q_{\nu})\Pi(q^2).
\end{equation}

Then we get the contribution of massless fermion to the vacuum polarization scalar
\begin{equation}
\Pi(q^2) = \frac{N_f}{8\,q}.
\end{equation}

The result is shown in Figure 7
\begin{figure}[h]
\includegraphics[width = 100mm]{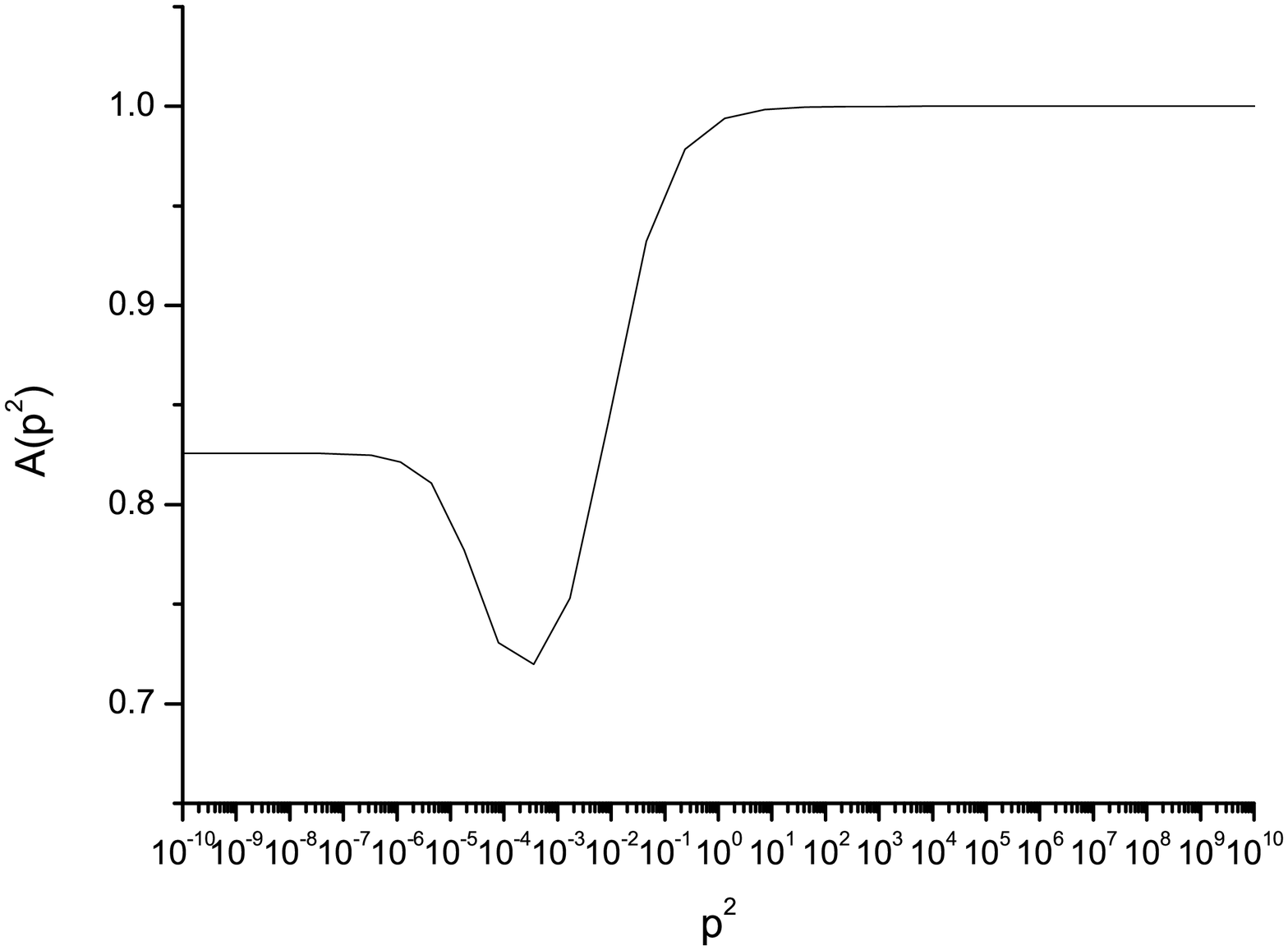}
\includegraphics[width = 100mm]{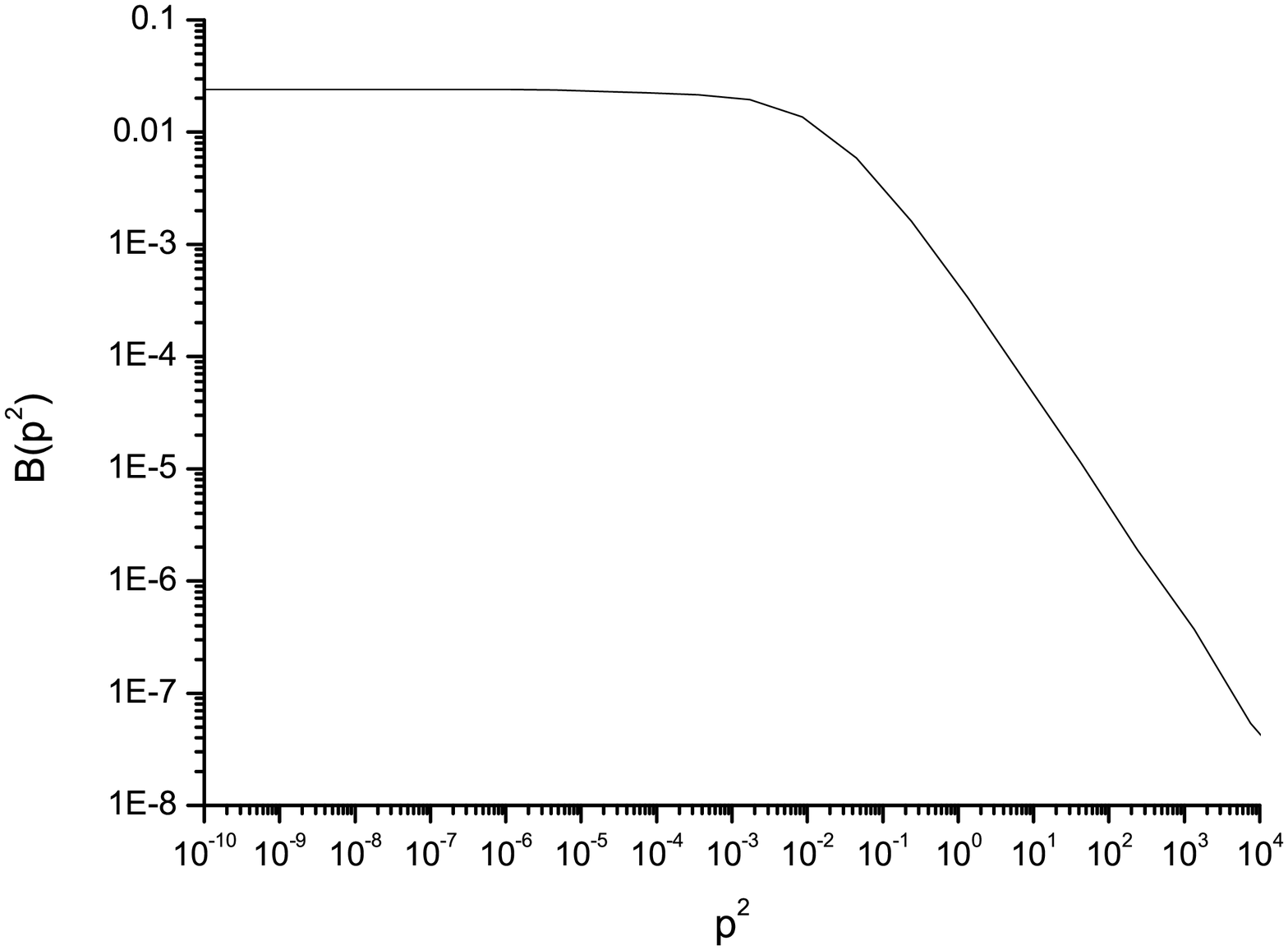}
\caption{The behavior of $A(p^2)$ and $B(p^2)$ with $N_f$ = 1 in the one-loop approximation.}
\end{figure}
\newpage
\begin{figure}[h]
\centering
\begin{fmffile}{6018}
\parbox{30mm}{
\begin{fmfgraph*}(30,10)
\fmfleft{i}
\fmfright{o}
\fmfv{d.sh=circle,d.f=hatched,d.si=.3w}{v}
\fmf{photon}{i,v,o}
\end{fmfgraph*}
}
$\approx$
\parbox{30mm}{\begin{fmfgraph*}(30,15)%\fmfkeep{fermion}
\fmfleft{i} \fmfright{o} \fmf{photon}{i,v1} \fmf{photon}{v2,o}
\fmf{fermion,right,tension=.55}{v1,v2,v1}
\end{fmfgraph*}}
\end{fmffile}
\caption{One-loop contribution to photon polarization.}
\end{figure}
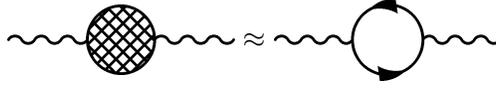
\subsection{Minimal BC vertex}
In the approximation of the minimal BC vertex, we only assume the first term in the BC vertex. This strategy introduces the fermion wavefucntion renormalization dependence in the vertex, while avoid the complication of singular structure in BC vertex.
\begin{figure}[h]
\includegraphics[width = 100mm]{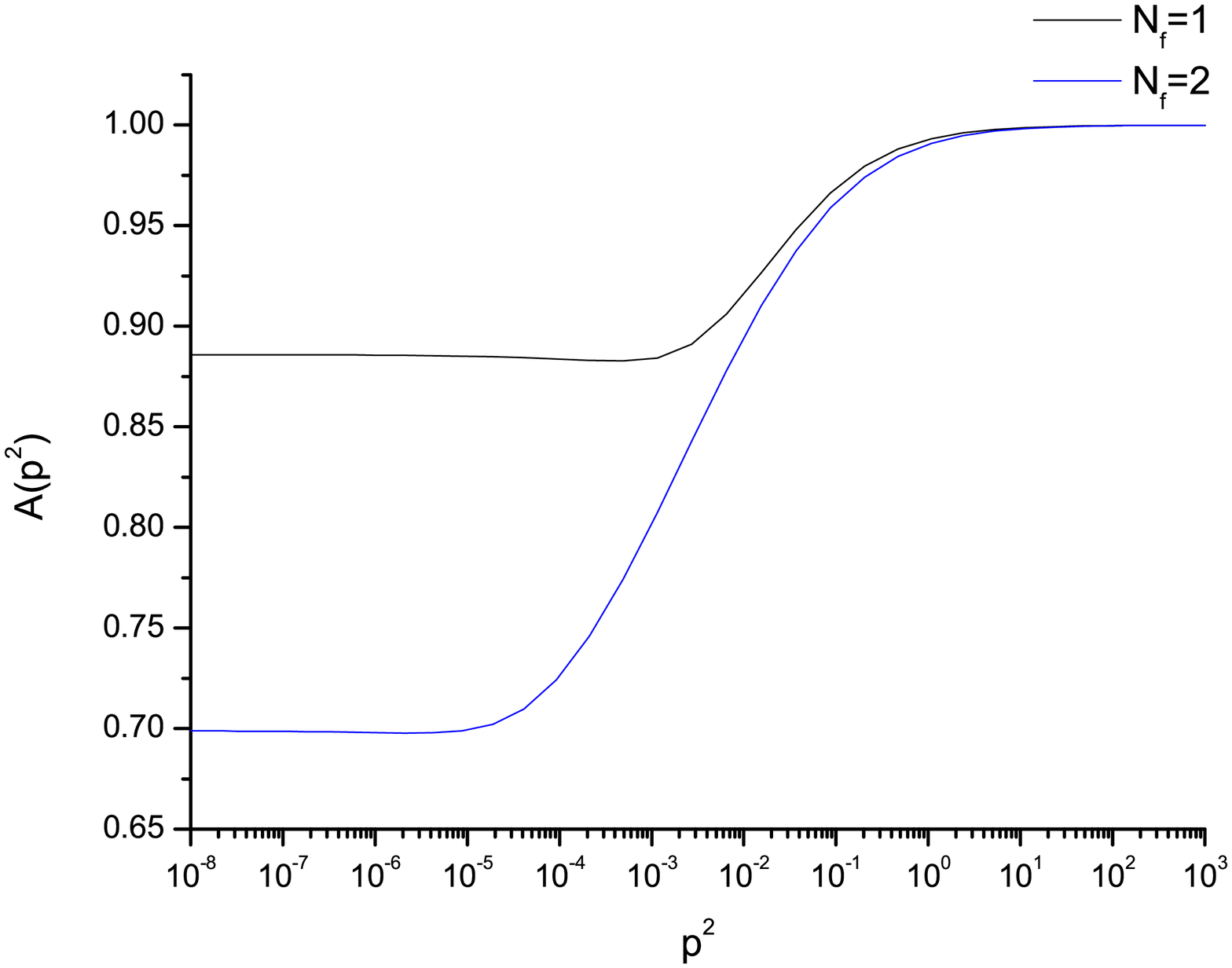}
\includegraphics[width = 100mm]{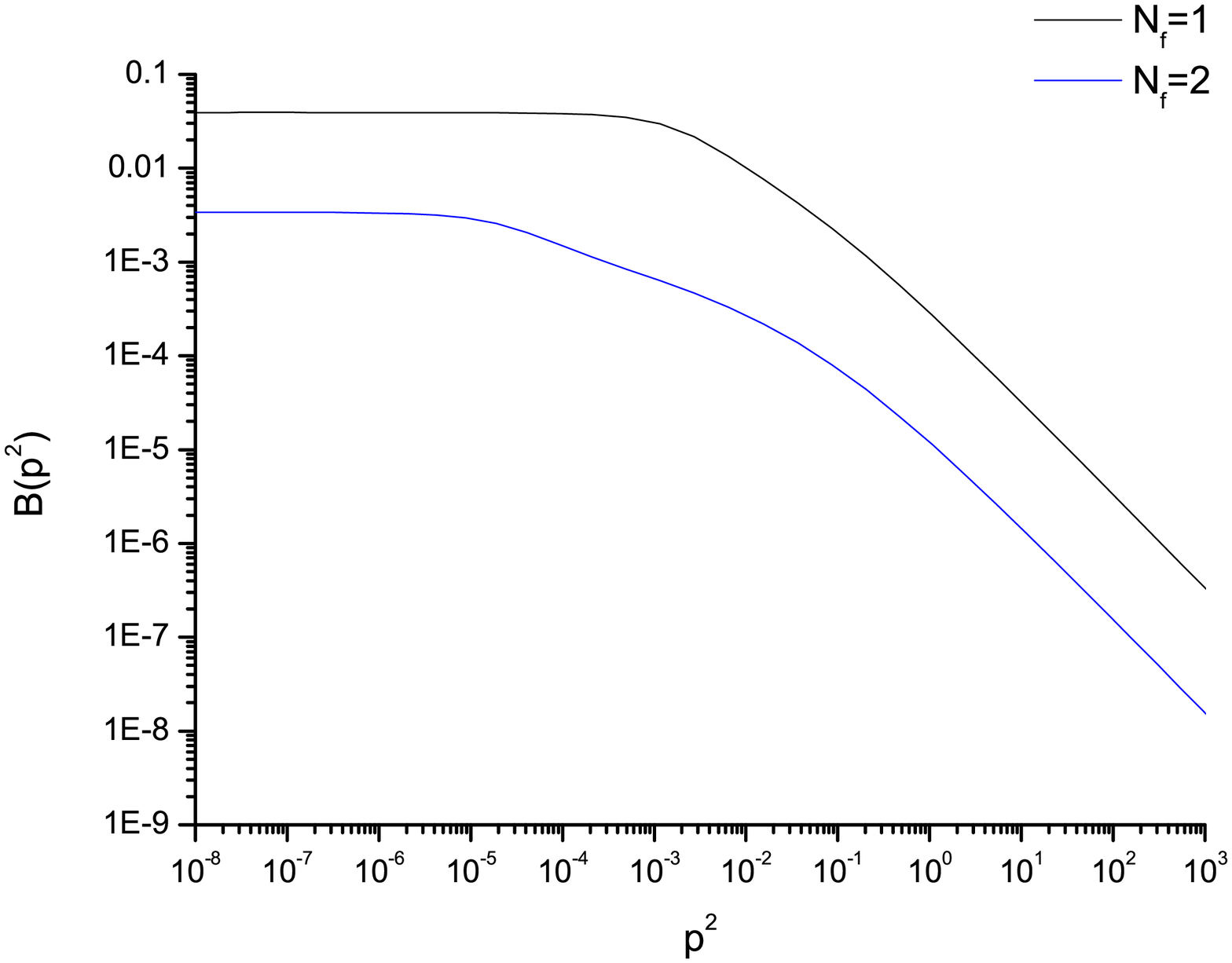}
\end{figure}
%\newpage
\begin{figure}[h]
\includegraphics[width = 100mm]{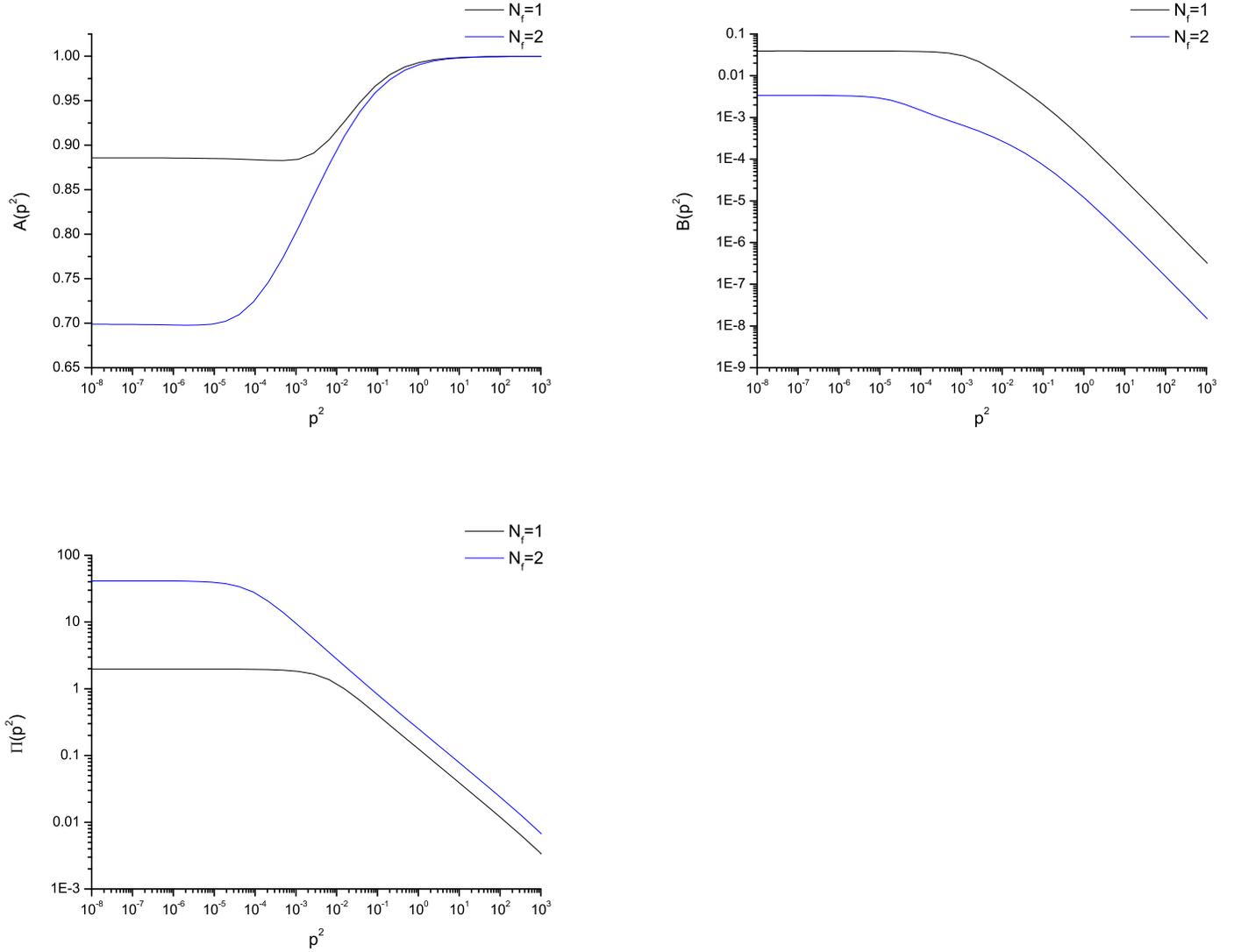}
\caption{The behavior of $A(p^2)$, $B(p^2)$ and $\Pi(p^2)$ in the minimal BC vertex.}
\end{figure}
\subsection{The complete BC vertex}
\subsubsection{The chirally broken phase}
The improved Newton's method is applied to the calculation of $A(p^2), B(p^2)$ and $\Pi(q^2)$. The iteration is extremely sensitive to initial values. The initial valued can be set to the results of the minimal BC vertex. The Nambu solution is shown in Figure 10.
\newpage
\begin{figure}[h]
\includegraphics[width = 100mm]{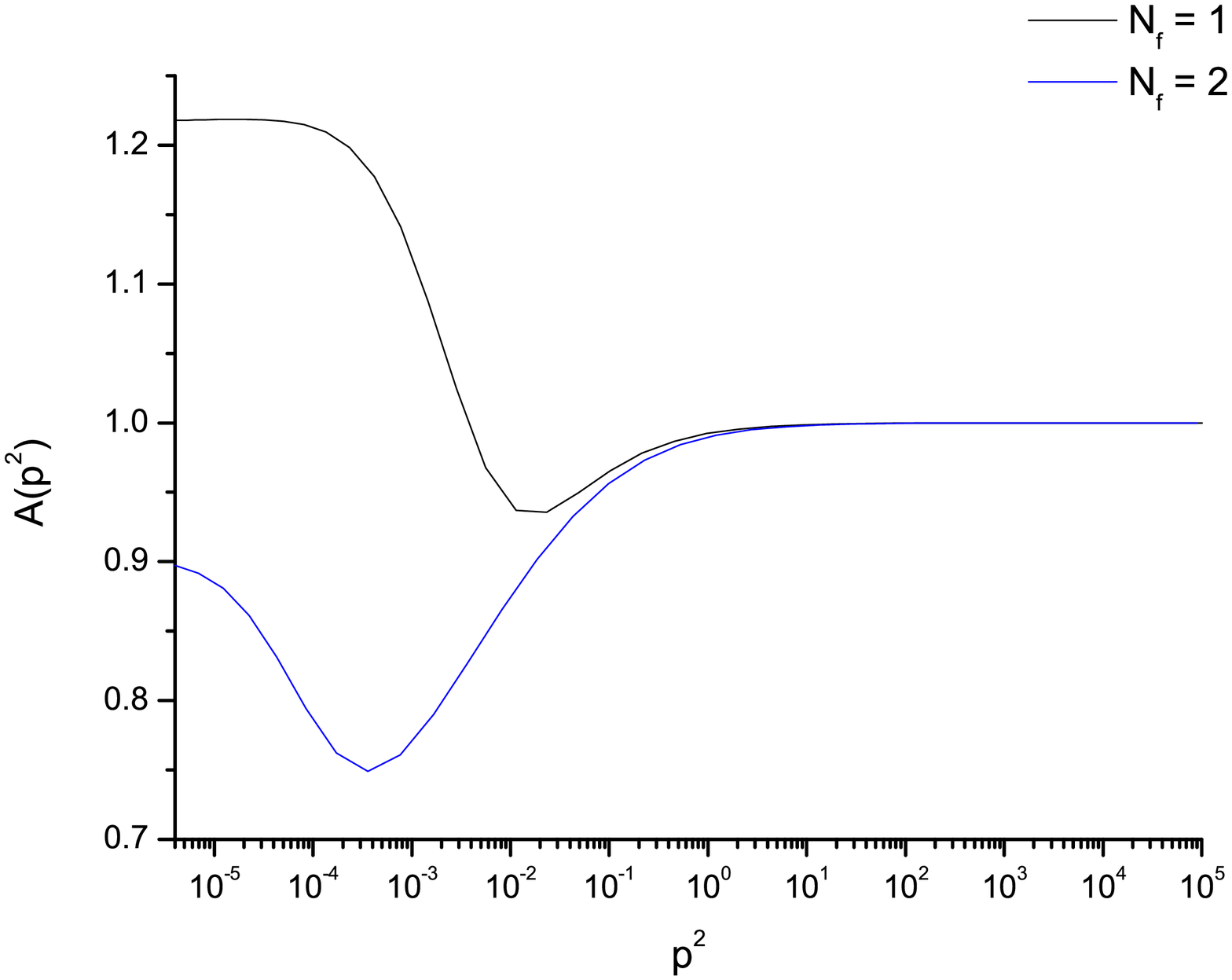}
\includegraphics[width = 100mm]{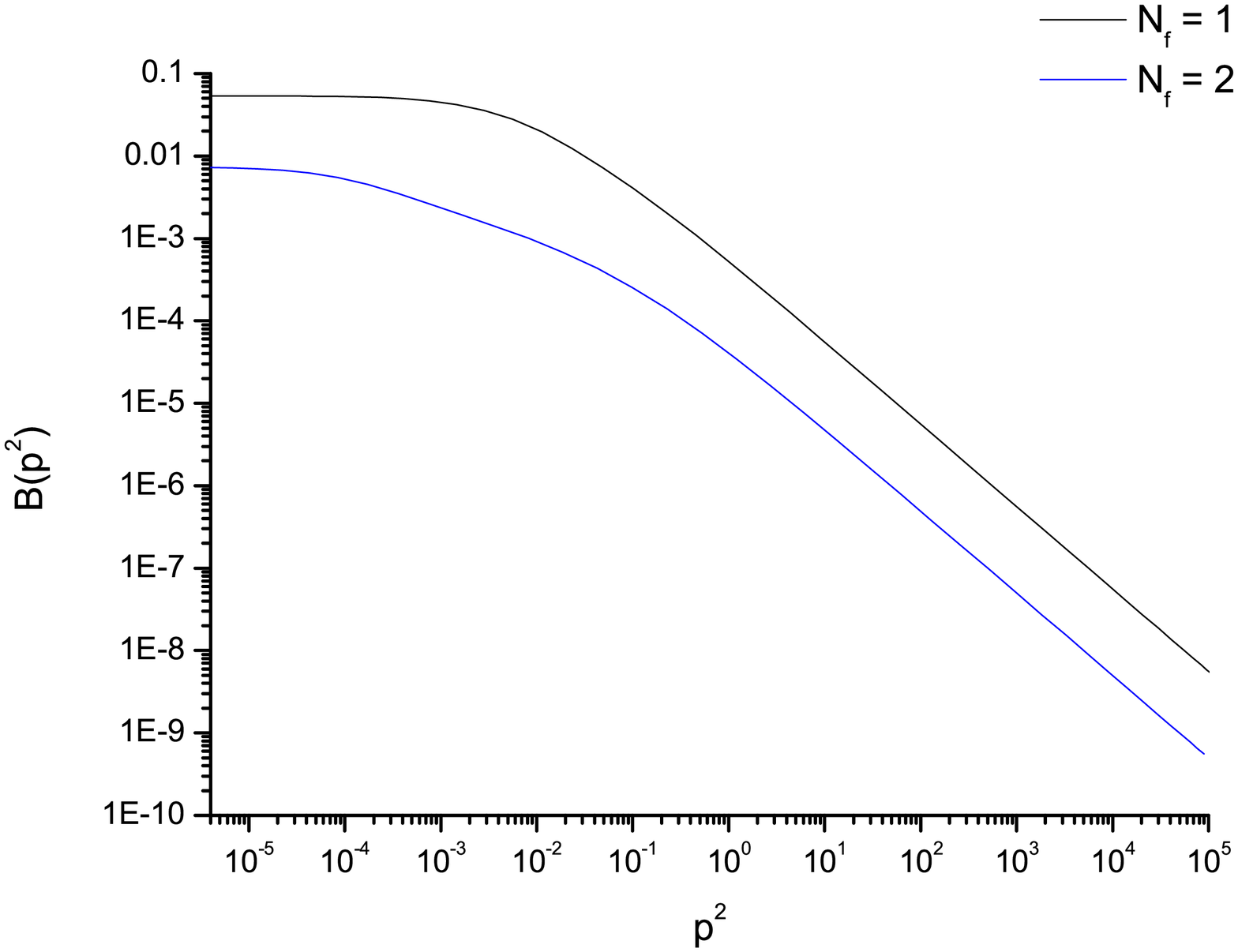}
%\includegraphics[width = 100mm]{BC_C}
%\caption{The behavior of $A(p^2)$, $B(p^2)$ and $\Pi{}$ in the BC vertex}
\end{figure}
\begin{figure}[h]
\includegraphics[width = 100mm]{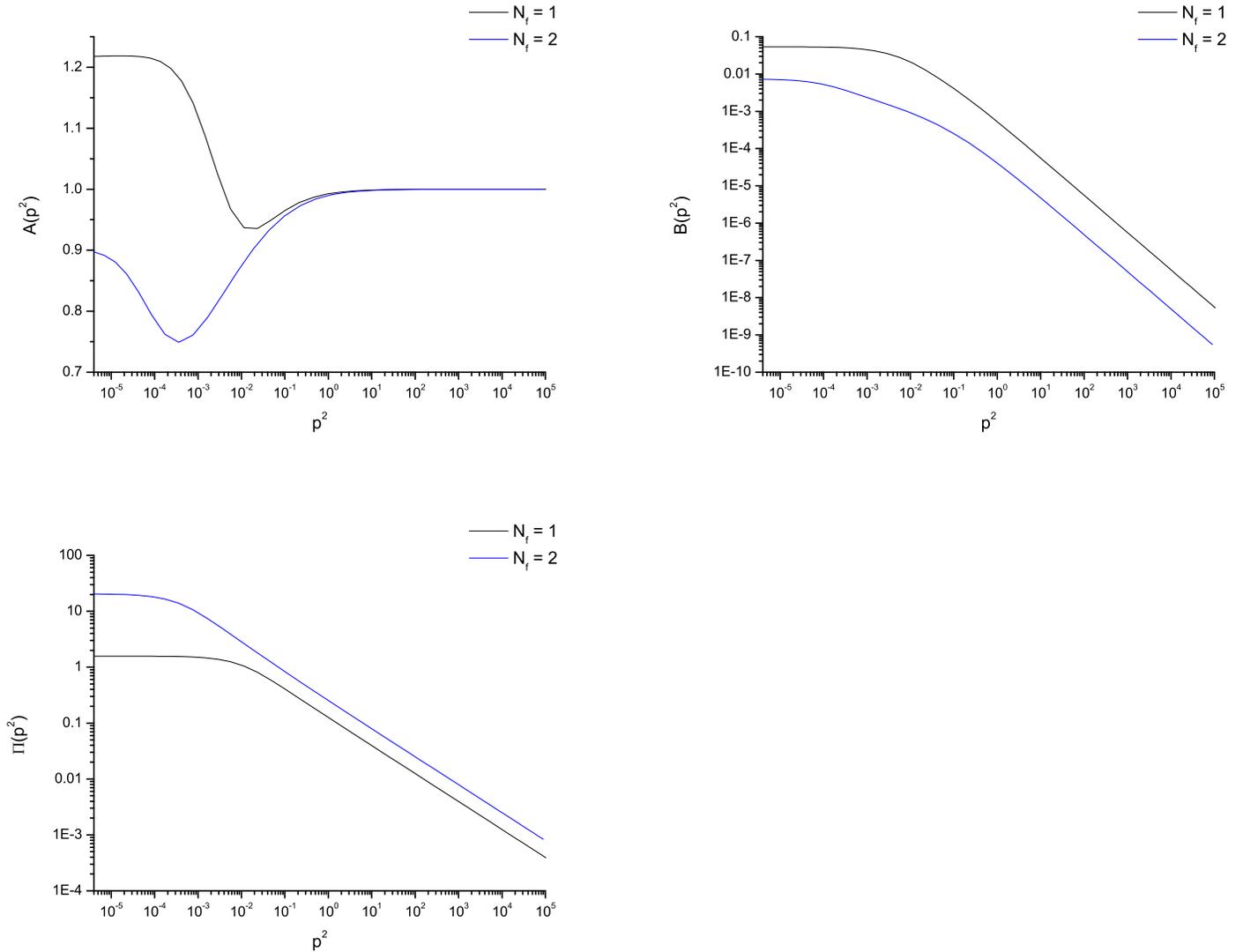}
\caption{The behavior of $A(p^2)$, $B(p^2)$ and $\Pi(p^2)$ in the BC vertex.}
\end{figure}
The above calculations show that the behaviors of dressing functions depends critically on truncation schemes. Comparison between the quenched case and the one-loop case shows that the contribution of one-loop vacuum polarization changes the behavior of $A(p^2)$ significantly, while having little bearings on the behavior of $B(p^2)$. After introducing wave-function renormalization in the vertex, although the behavior of $A(p^2)$ is modified slightly, $B(p^2)$ remains almost the same. The vacuum polarization $\Pi(p^2)$ has similar behavior to that of $B(p^2)$. When the complete BC vertex is considered, again the shape of $A(p^2)$ is changed drastically, while $B(p^2)$ and $\Pi(p^2)$ show little change. We thus reach the conclusion that the behaviors of $B(p^2)$ and $\Pi(p^2)$-which represent contributions to the fermion and photon self-energy, respectively-are insensitive to truncation schemes, while $A(p^2)$-which contributes to the wave-function renormalization-demonstrates close connection to the form of the fermion-photon vertex. We also find a model-independent result that although $A(p^2)$ shows different behaviors in different approximations, it approaches its asymptotic form at $p^2\approx0$ and for $B(p^2)$ and $\Pi(p^2)$ this happens at roughly $p^2\approx10^{-2}$.
\subsubsection{The symmetric phase}
The chirally symmetric phase (often referred to as Wigner phase, and the corresponding solution of the gap equation is called the Wigner solution) of the DSEs is necessary in the discussion of phase transition problems in QED$_3$ and QCD~\cite{z05,q11,j12,w12,c13,c14}. This solution can be obtained by setting $B(p^2)\equiv0$ and it has been found by the authors of Ref.~\cite{qy} in the bare and BC1 vertices. However, no solution in the complete BC vertex has been reported before. We therefore make an attempt to calculate the Wigner solution in BC vertex adopting the method stated above.

The power-law form $A(p^2) = c\, p^{2\kappa}$ is generally assume in the infrared region. In the case of bare and minimal BC vertex (or BC1), self-consistent solutions satisfying DSEs have been found and are related to flavor number by\cite{ToyModel}
\begin{equation}
\kappa_{bare} = \frac{0.135}{N_f} + \frac{0.090}{N_f^2} + O(\frac{1}{N^3_f}),
\end{equation}
and for BC1 vertex, two solutions have been found
\begin{equation}
\kappa_{BC1}=\frac{0.115}{N_f} + \frac{0.044}{N^2_f} + O(\frac{1}{N_f^3}),
\end{equation}
\begin{equation}
\kappa_{BC1} = 0.5 - \frac{0.050}{N_f}-\frac{0.006}{N^2_f}-\frac{0.028}{N_f^3}.
\end{equation}

However, the complete BC vertex doesn't admit self-consistent power-law form of solutions. Thus we take the empirical extrapolation with $\kappa = 0.12$, which can be continuously matched to the numerical solutions with the condition
\begin{equation}
\frac{A(p^2)}{p^{2 \kappa}} = \frac{A(\varepsilon^2)}{\varepsilon^{2\kappa}},
\end{equation}
where $\varepsilon$ is the infrared cutoff.

In the ultraviolet region, the solutions restore to the free form. Then we have $A(p)=1, \Pi(q) \propto 1/q$. The result is shown in Figure 11.
\begin{figure}[h!]
\includegraphics[width = 93mm]{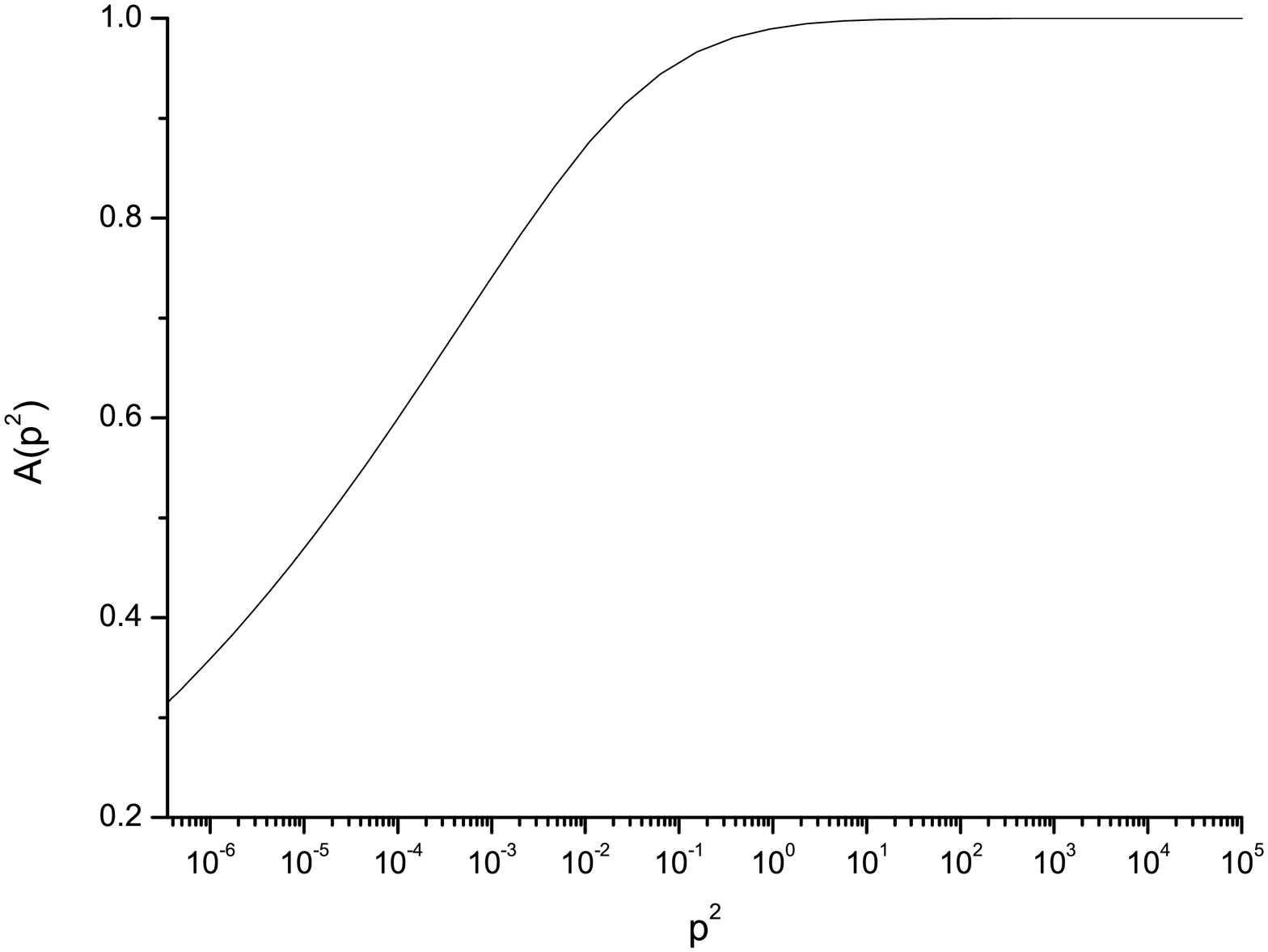}
\includegraphics[width = 93mm]{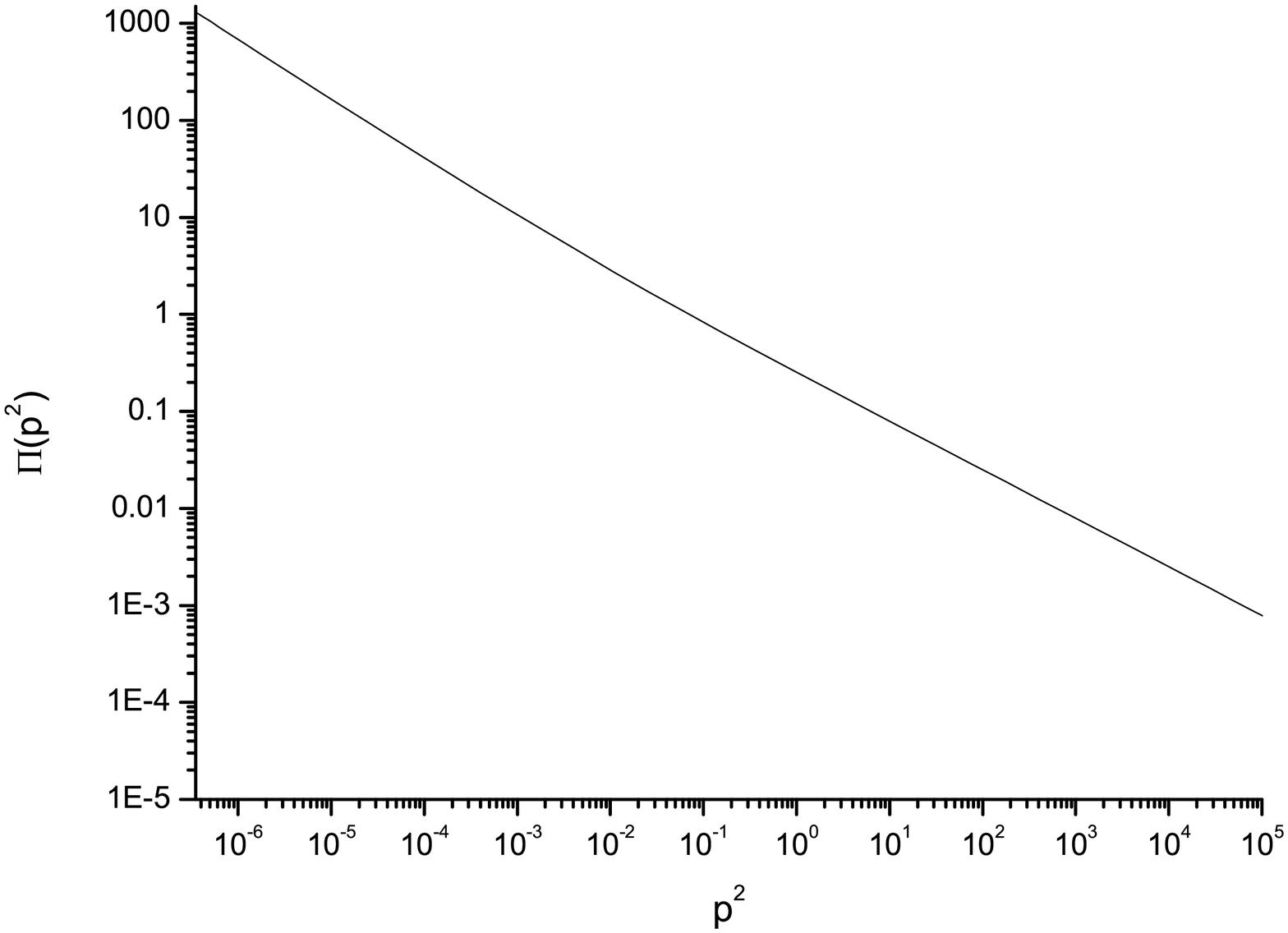}
\caption{The Wigner solution of $A(p^2)$ and $\Pi(p^2)$ with $N_f$ = 2 in the BC vertex.}
\end{figure}

Due to the complexity of the problem, this solution was not reported in any known literature, and by comparison with the results presented in \cite{qy}, we find that when adopting the complete BC vertex, $A(p^2)$ approaches $0$ more rapidly as $p\rightarrow 0$, while $\Pi(p^2)$ remains insensitive to truncation schemes.
\section{Conclusions}
An improvement is made in the calculation of dressing functions in QED$_3$. This method has shown better convergence than other existing methods, especially when applied to the calculation of vertices with singular structure. This method can be adopted in the calculation of dressing function in more flavors, and thus determine the flavor dependence of quantities such and fermion-antifermion condensate.

Furthermore, attempts have been made to search for Wigner solution of DSEs in BC vertex. Similarly, we can further explore the flavor dependence of Nambu and Wigner solution of DSEs on the basis of this method, the results can be applied to problems such as determining the critical point of phase transition.
\section*{Acknowledgments}
This work is supported in part by the National Natural Science Foundation of China (under Grant Nos. 11275097 and 11475085).

\begin{appendices}
\section*{Appendix A. Trace Techniques}
In three dimensional Euclidean space
\begin{equation*}
$tr$(I) = 4,
\end{equation*}
\begin{equation*}
$tr$(\gamma^{\mu}\gamma^{\nu}) = 4\delta^{\mu\nu},
\end{equation*}
\begin{equation*}
$tr$(p\!\!\!/_1\,p\!\!\!/_2) = 4 p_1\cdot p_2,
\end{equation*}
\begin{equation*}
$tr$(p\!\!\!/_1\,p\!\!\!/_2\,p\!\!\!/_3\,p\!\!\!/_4) = 4[(p_1\cdot p_2)(p_3\cdot p_4)-(p_1\cdot p_3)(p_2\cdot p_4) + (p_1\cdot p_4)(p_2 \cdot p_3)],
\end{equation*}
\begin{equation*}
$tr$(p\!\!\!/_1\,p\!\!\!/_2,...p\!\!\!/_n) = 0, \ \ \ \ $when n is odd number$,
\end{equation*}
where $\gamma_{\mu}$ is 4-dimensional matrix satisfying
\begin{equation*}
\{\gamma_{\mu},\gamma_{\nu}\} = 2\, \delta_{\mu\nu}.
\end{equation*}
\end{appendices}

\end{document}